\documentclass[aip,reprint]{revtex4-1}

\usepackage{color}
\usepackage{graphicx}
\usepackage{amsmath}
\usepackage{subcaption}

\begin{document}
	\title{Magnetized ICF Implosions: Scaling of Temperature and Yield Enhancement}
	
	\author{C. A. Walsh}
	\affiliation{Lawrence Livermore National Laboratory, USA}
	
	\author{S. O'Neill}
		\affiliation{Imperial College, London, UK}
	\author{J. P. Chittenden}
		\affiliation{Imperial College, London, UK}
	\author{A. J. Crilly}
		\affiliation{Imperial College, London, UK}
	\author{B. Appelbe}
		\affiliation{Imperial College, London, UK}
	\author{D. J. Strozzi}
		\affiliation{Lawrence Livermore National Laboratory, USA}
	\author{D. Ho}
		\affiliation{Lawrence Livermore National Laboratory, USA}
	\author{H. Sio}
		\affiliation{Lawrence Livermore National Laboratory, USA}
	\author{B. Pollock}
		\affiliation{Lawrence Livermore National Laboratory, USA}
	\author{L. Divol}
	\affiliation{Lawrence Livermore National Laboratory, USA}
	\author{E. Hartouni}
	\affiliation{Lawrence Livermore National Laboratory, USA}
	\author{M. Rosen}
	\affiliation{Lawrence Livermore National Laboratory, USA}
	\author{B. G. Logan}
	\affiliation{Lawrence Livermore National Laboratory, USA}
	\author{J. D. Moody}
		\affiliation{Lawrence Livermore National Laboratory, USA}
	
	\email{walsh34@llnl.gov}
	
	\date{\today}

	\begin{abstract}
		This paper investigates the impact of an applied magnetic field on the yield and hot-spot temperature of inertial confinement fusion implosions. A scaling of temperature amplification due to magnetization is shown to be in agreement with unperturbed 2-D extended-magnetohydrodynamic simulations. A perfectly spherical hot-spot with an axial magnetic field is predicted to have a maximum temperature amplification of 37\%. However, elongation of the hot-spot along field lines raises this value by decreasing the hot-spot surface area along magnetic field lines. A scaling for yield amplification predicts that a magnetic field has the greatest benefit for low temperature implosions; this is in agreement with simplified 1-D simulations, but not 2-D simulations where the hot-spot pressure can be significantly reduced by heat-flow anisotropy. Simulations including a P2 drive asymmetry then show that the magnetized yield is a maximum when the capsule drive corrects the hot-spot shape to be round at neutron bang-time. The benefit of an applied field increases when the implosion is more highly perturbed. Increasing the magnetic field strength past the value required to magnetize the electrons is beneficial due to additional suppression of perturbations by magnetic tension.
	\end{abstract}
	\maketitle
	
	\section{Introduction}
	
	Magnetic fields applied to inertial confinement fusion (ICF) implosions compress with the plasma \cite{gotchev2009,knauer2010}, reaching strengths that can magnetize electron thermal conduction losses from the hot fuel. This was first demonstrated on the OMEGA Laser Facility, where an axial 8T magnetic field increased the hot-spot temperature and yield by 15\% and 30\% respectively. Larger magnetic fields are anticipated to have a greater impact on the target performance \cite{walsh2019}, with some calculations showing a relaxation of the ignition cliff when thermal conduction is magnetized \cite{perkins2017,perkins2013}.
	
	In addition to improvements in bulk target performance, an applied field fundamentally changes perturbation growth. Tension in the magnetic field lines through the Lorentz force can stabilize the Rayleigh-Taylor instability (RTI) \cite{chandrasekhar1962}; 2-D simulations have demonstrated severe reductions in growth rates due to this phenomenon \cite{srinivasan2013,perkins2017,PhysRevE.104.L023201} (although recent 3-D simulations demonstrated that the 2-D RTI suppression is greatly exaggerated \cite{walsh2021magnetized}). In addition, suppressing thermal conduction into perturbations reduces ablative stabilization and can result in greater perturbation growth during hot-spot stagnation \cite{walsh2019,walsh2021magnetized}. In direct-drive implosions a magnetic field can also result in greater drive-phase asymmetries by suppressing non-radial heat-flow in the conduction zone \cite{walsh2020a,matsuo2017}.
	
	This paper demonstrates a simple scaling of hot-spot temperature and yield amplification due to an applied magnetic field. This scaling is shown to be applicable to a wide range of ICF designs, including both directly and indirectly driven systems and implosions of both spherical and cylindrical geometries. The changing impact of a magnetic field when low-mode drive asymmetries and higher mode perturbations are present is then explored.
	
	In the presence of a magnetic field, electron thermal conduction becomes anisotropic:
	
	\begin{equation}
	\underline{q} = -\kappa_{\parallel} \left(\underline{\hat{b}} \cdot \nabla T_e \right) \underline{\hat{b}} -\kappa_{\bot} \underline{\hat{b}} \times \left(\nabla T_e \times \underline{\hat{b}}\right)  - \kappa_{\wedge} \underline{\hat{b}} \times \nabla T_e\label{eq:heatflow}
	\end{equation}
	
	Where $\underline{\hat{b}}$ is the magnetic field unit vector. The first term on the right ($\kappa_{\parallel}$) is heat-flow along magnetic field lines, which is unchanged by magnetization. The second term ($\kappa_{\bot}$) is heat-flow perpendicular to magnetic field lines, which becomes suppressed as the electron magnetization increases. The magnetization is characterized by the Hall Parameter, $\omega_e \tau_e$ (the product of the electron gyrofrequency and the electron-ion collision time), which scales as:
	
	\begin{equation}
		\omega_e \tau_e \sim \frac{T_e^{3/2} | \underline{B}|}{\rho Z_{eff}} \label{eq:wt}
	\end{equation}
	
	where $Z_{eff} = <Z^2>/<Z>$ is the effective ionization averaged over ion species. Equation \ref{eq:wt} shows that a low density and high temperature plasma is more easily magnetized than a high density, low temperature plasma. A DT plasma with $\omega_e \tau_e = 1$ will have perpendicular thermal conductivities suppressed by 69\% ($\kappa_{\bot}/\kappa_{\parallel} = 0.31$) \cite{epperlein1986}. For $\omega_e \tau_e = 10$ this is reduced to $\kappa_{\bot}/\kappa_{\parallel} < 0.01$.
	
	The final term in equation \ref{eq:heatflow} is the Righi-Leduc heat-flow, which represents the heat-flow deflected by the magnetic field, transferring heat perpendicular to the temperature gradient. For a DT plasma this effect peaks for $\omega_e \tau_e \approx 0.6$, giving $\kappa_{\wedge}/\kappa_{\parallel} = 0.44$. Righi-Leduc heat-flow is typically ignored in magnetized ICF implosions, for two reasons. Firstly, the Hall Parameter of the compressed fuel is typically much greater than 1, resulting in $\kappa_{\wedge}$ being small. Secondly, implosions are most frequently modeled in 2-D \cite{davies2015,davies2017,perkins2013,perkins2017,slutz2010,slutz2012,slutz2016,slutz2018,johzaki2016,2021}, where $\underline{\hat{b}} \times \nabla T_e$ is mostly out of the simulation plane and has no effect. However, 3-D simulations of the magnetized ablative-RTI have found that Righi-Leduc can seed higher mode perturbations when the plasma is moderately magnetized \cite{walsh2021magnetized}.
	
	While the heat-flow in equation \ref{eq:heatflow} is anisotropic, it is useful to define an effective thermal conductivity, $\kappa_{eff}$, for characterizing magnetized implosions. $\kappa_{eff}$ is defined such that an unmagnetized implosion has $\kappa_{eff}=1$. The effective thermal conductivity has different definitions in cylindrical and spherical geometries.
	
	Magnetized cylindrical implosions have been conducted on the OMEGA Laser Facility, both with \cite{davies2017,barnak2017} and without \cite{hansen2020} laser preheat. If the axial length-scale is assumed to be much greater than the radial, the heat-flow is everywhere perpendicular to an applied axial magnetic field. This orientation is optimal for magnetization, allowing for full suppression of electron heat-flow. Therefore, for cylindrical implosions, $\kappa_{eff} = \kappa_{\bot}/\kappa_{\parallel}$. For the purposes of this paper, $\kappa_{\bot}$ is calculated using the burn-averaged $\omega_e \tau_e$.
	
	Magnetized spherical implosions without applied perturbations are inherently two dimensional; an axial field is normal to the capsule surface at the pole and in the plane of the surface at the waist. If the plasma is completely magnetized ($\omega_e \tau_e \gg 1$) then the heat-flow will be suppressed in 2 out of the 3 possible dimensions. Therefore, for spherical implosions, $\kappa_{eff} = 1/3 + 2\kappa_{\bot}/3\kappa_{\parallel}$ \cite{chang2011,ho2016}. In reality the applied axial magnetic field bends during the implosion, resulting in a non-axial component that reduces the magnetic field effect.
	
	A scaling for the hot-spot temperature with $\kappa_{eff}$ can be simply derived from the change in temperature due to a heat-flow divergence \cite{2018}:
	
	\begin{equation}
		T_{hs}^{7/2} \sim \frac{R_{hs}^2 P_{hs}}{t_{stag} \kappa_{eff}}
	\end{equation}
	
	Where $T_{hs}$ is the hot-spot temperature (assuming the electrons and ions are fully equilibrated), $R_{hs}$ is the hot-spot radius, $P_{hs}$ is the hot-spot pressure and $t_{stag}$ is the stagnation time-scale. A Spitzer thermal conductivity has been assumed \cite{spitzer1953}. By assuming that the hot-spot radius, pressure and stagnation time are not affected by the magnetic field, the temperature scales as $T \sim \kappa_{eff}^{-2/7}$ A more thorough derivation and discussion will be given in section \ref{sec:temp_amp}. This scaling, as well as the subsequent scaling for fusion yield, will be compared with two types of simulations. Firstly, 1-D calculations using a prescribed $\kappa_{eff}$ are conducted. These do not include any MHD physics; instead the thermal conductivity in the fuel is artificially reduced. By keeping the simulations 1-D there is no impact from thermal conduction anisotropy on the implosion shape. Subsequently, the scalings are tested against full 2-D extended-MHD simulations, where an applied field strength is prescribed and the effective thermal conductivity is calculated using the burn averaged electron magnetization.
	
	The simulations in this paper use the Gorgon extended-MHD code \cite{ciardi2007,chittenden2009,walsh2017}. Heat-flow in Gorgon is treated anisotropically, using a scheme that is designed to minimize numerical diffusion perpendicular to the field lines \cite{sharma2007,walsh2018a}. The magnetic transport includes bulk plasma advection, resistive diffusion, Nernst and cross-gradient-Nernst \cite{walsh2020}, as well as Biermann Battery magnetic field generation \cite{walsh2017,walsh2021a,PhysRevLett.125.145001,campbell2021measuring}. The magnetic transport algorithm uses updated magnetic transport coefficients \cite{sadler2021,davies2021} that have been shown to reduce magnetic field twisting in pre-magnetized implosions \cite{2021}. Gorgon magnetic flux compression simulations compare favorably with experiments \cite{gotchev2009,10.1088/1361-6587/ac3f25}, increasing confidence in the bulk and Nernst advection of magnetic field. 
	
	A series of ICF designs are used in this paper in order to demonstrate the applicability of the proposed temperature and yield scalings. 
	
	First, a layered DT implosion N170601 is used, which is a cryogenic experiment from the HDC campaign \cite{clark2019}. This shot has moderate $\alpha$-heating and was the first to exceed $10^{16}$ neutrons. The scaling proposed in this paper is not applicable in the ignition regime, where small changes in implosion quality can result in large changes in yield \cite{Tong_2019}. Therefore, no $\alpha$-heating is included in these simulations. For the impact of magnetization on propagating thermonuclear burn, see \cite{appelbe2021}.
	
	A room temperature (warm) HDC indirect-drive design for the National Ignition Facility (NIF) is also employed, which uses a gaseous deuterium fuel (4kg/m$3$) and a 1.1MJ laser pulse. This design is similar to the symmetry capsules (symcaps) used in the big-foot campaign \cite{casey2018,thomas2020}. While this design does not produce a high yield, it is the starting point for demonstrating the benefits of magnetization on NIF \cite{moody2021,moody2021a}. If these proposed experiments on NIF are successful, it will motivate the engineering advancements needed to magnetize a high yield layered DT capsule. 
	
	All indirect-drive drive simulations in this paper do not incorporate the hohlraum. The simulations use a prescribed frequency-dependent source that has been calculated using hohlraum simulations; it is assumed that the magnetic field does not change the X-ray drive. For hohlraum calculations including applied magnetic fields, see \cite{strozzi2015,montgomery2015,strozzi}
	
	A cryogenic layered DT OMEGA implosion design is also simulated \cite{PhysRevLett.100.185006,crilly2020}, demonstrating that the magnetization effect is applicable to scaled-down direct-drive designs as well. These implosions are investigated without any perturbation sources, due to the added complications of instability growth in a magnetized direct-drive ablation front \cite{walsh2020a,PhysRevLett.127.165001,bose2021}. 
	
	A cylindrical design is taken from the magnetized implosions conducted on the OMEGA Laser Facility without any fuel preheat \cite{hansen2020,10.1088/1361-6587/ac3f25}. While this design compares well with theory for the simplified 1-D simulations with prescribed effective thermal conductivity, in practice the magnetic pressure is significant in these implosions \cite{hansen2020}, and is not accounted for in the theoretical scaling.
	
	The structure of this paper is as follows. Section \ref{sec:temp_amp} outlines the temperature amplification scaling and shows that it compares favorably with symmetrically-driven simulations. Section \ref{sec:Y_amp} discusses the expected impact of the temperature amplification on neutron yield assuming that the hot-spot pressure remains unchanged by magnetization. As the magnetized hot-spots are found to elongate along the magnetic field lines (reducing the hot-spot pressure), section \ref{sec:shape} looks at the performance benefits of correcting for that inherent asymmetry; once the shape is corrected the agreement with theoretical yield amplification from section \ref{sec:Y_amp} is improved. The final two sections then focus on the temperature and yield enhancements for asymmetric implosions. Section \ref{sec:P4_shape} studies the axi-symmetric P4 asymmetry, while section \ref{sec:pert} uses higher mode perturbations; the benefit of a magnetic field is increased for perturbed implosions in both cases.
	
	\section{Temperature Amplification \label{sec:temp_amp}}
	
	\begin{figure}
		\centering
		\includegraphics[width=0.5\textwidth]{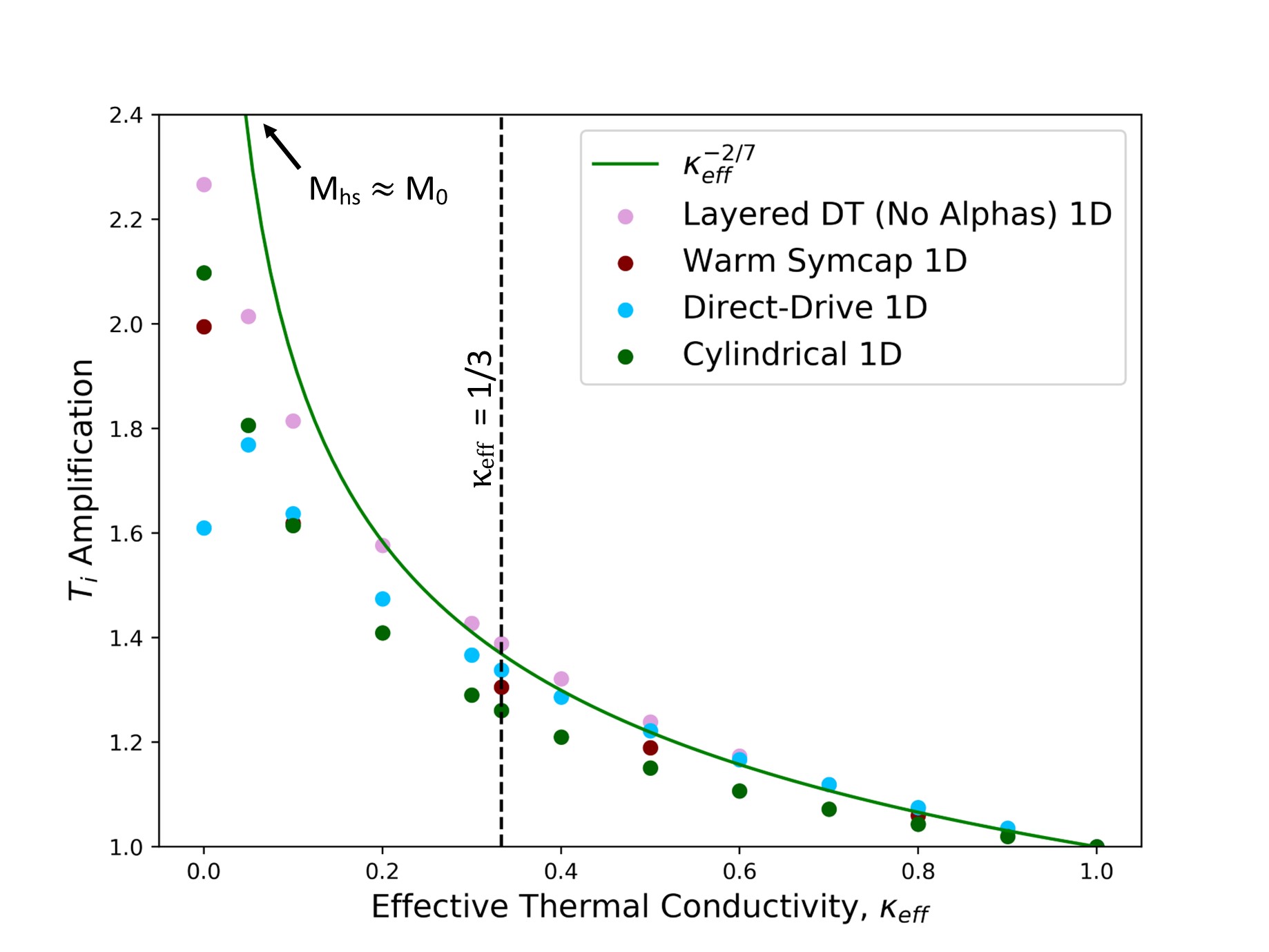}
		\caption{ \label{fig:Ti_kappamod} Temperature amplification by suppression of fuel electron thermal conductivity for several different target designs. Here the effective thermal conductivity, $\kappa_{eff}$ is prescribed, and is not based on any MHD description. All simulations are 1-D, which is not possible once MHD physics is self-consistently included. $T_i$ is a burn averaged quantity.}
	\end{figure}
	
	\begin{figure}
		\centering
		\includegraphics[width=0.5\textwidth]{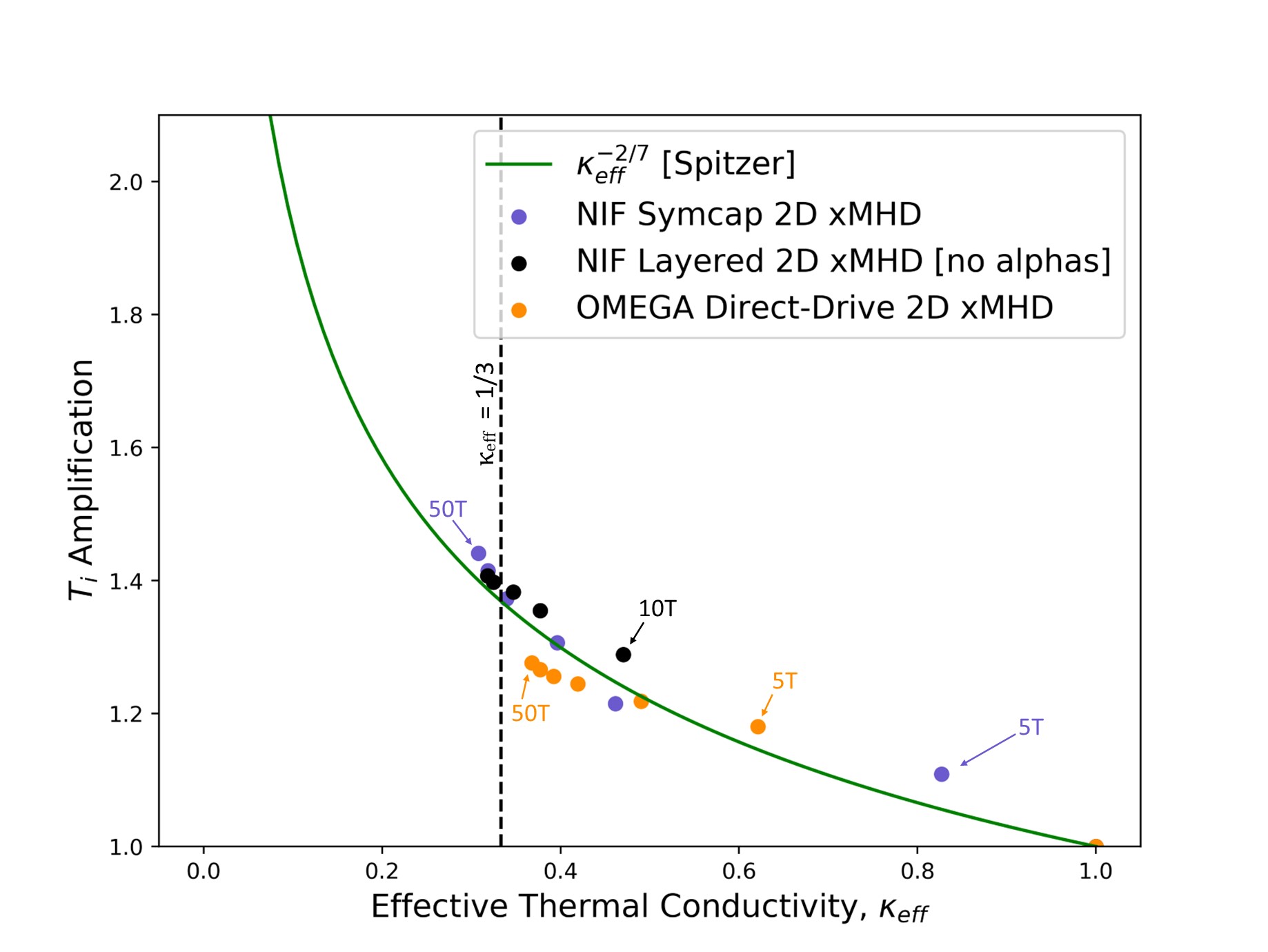}
		\caption{ Simulated temperature amplification from an applied magnetic field for 3 target designs. The effective thermal conductivity is calculated by using the burn-averaged Hall Parameter and the hot-spot shape. The simulations are in two dimensions, allowing for natural shape asymmetries to develop due to the anisotropic thermal conduction. $T_i$ is a burn averaged quantity. \label{fig:Ti_xMHD} }
	\end{figure}
	
	This section explores how the temperature of an ICF hot-spot changes with varying thermal conductivity. The magnetic field is assumed to only affect the plasma during stagnation. A spherical hot-spot is used, although the resultant scaling is identical in cylindrical geometry.
	
	At the beginning of stagnation, when the first shock has converged on axis and is expanding out through the hot-spot, the hot-spot is taken to have mass $M_0$. Ablation of cold fuel into the hot-spot at velocity $V_{abl}$ over the stagnation time-scale $\tau_{stag}$ gives a hot-spot mass:
	
	\begin{equation}
		M_{hs} = M_0 + 4\pi R_{hs}^2 \tau_{stag} \rho_{sh} V_{abl} \label{eq:M_hs}
	\end{equation}
	
	Where $R_{hs}$ is the hot-spot radius and $\rho_{sh}$ is the shell density. Mass ablation is induced by any process that heats the colder fuel at the hot-spot edge. Hot-spots are approximately isobaric, meaning that any transfer of heat results in the transport of mass in the opposite direction. A generic hot-spot power source $\Psi$ (in units of J/s) heats up enough cold fuel such that:
	
	\begin{equation}
		 V_{abl} = \frac{\Psi}{(T_{hs}-T_{sh})n_{i,sh} (1+Z) 4 \pi R_{hs}^2}
	\end{equation}
	where $T_{sh}$ and $n_{i,sh}$ are the shell temperature and ion number density. $T_{hs}$ is the hot-spot temperature. An ideal gas equation of state has been assumed. 
	
	Electron thermal conduction is the dominant source of power transfer between the hot-spot and shell in a typical ICF implosion \cite{betti2001,Tong_2019}. In this case the thermal power transfer is $\Psi_{q_e} = 4 \pi R_{hs}^2 \kappa_{eff} n_e T_{e} \tau_e \nabla T_e / m_e$. Approximating $\nabla T_e \approx (T_{hs}-T_{sh})/R_{hs}$ and using the hot-spot quantities to evaluate the bulk parameters gives the hot-spot ablation velocity:
	
	\begin{equation}
		V_{abl} = \frac{\kappa_{eff} a T_{hs}^{5/2}}{R_{hs} \rho_{sh} ln\Lambda} + V_{abl,other} \label{eq:V_abl}
	\end{equation}
	
	Where $a$ is a pre-factor that depends on the plasma composition and $ln\Lambda$ is the Coulomb logarithm. For a more thorough derivation of this relation, see \cite{betti2001}; in that case the pre-factor $a$ also depends on geometry. $V_{abl,other}$ is the ablation velocity originating from other power sources, e.g. radiation, $\alpha$-heating or ion thermal conduction. The impact of radiation will be discussed at the end of this section, but for now it is assumed that $V_{abl,other}=0$. Combining equations \ref{eq:M_hs} and \ref{eq:V_abl} gives:
	
	\begin{equation}
		(M_{hs}-M_0) \rho^{5/2} = \frac{4\pi R_{hs} \tau_{stag} \kappa_{eff} a P_{hs}^{5/2} m_i^{5/2}}{ln\Lambda (1+Z)^{5/2}} \label{eq:M_hs2}
	\end{equation}
	
	Where the temperature has been factored out by using the relation for the hot-spot pressure $P_{hs} =n_{i,hs}T_{hs} (1+Z)$.
	
	Next, a number of key assumptions are made that will be tested and discussed throughout this paper. First, the initial hot-spot mass is assumed to be much smaller than the final hot-spot mass ($M_0 \ll M_{hs}$); the validity of this assumption depends on the specific capsule design in question and decreases in validity as the mass ablation is more suppressed. 
	
	Secondly, it is assumed that only $\kappa_{eff}$ on the right hand side of equation \ref{eq:M_hs2} varies with applied field. For the hot-spot radius ($R_{hs}$) and stagnation time-scale ($\tau_{stag}$) this is found from simulations to be a good assumption. The invariance of the hot-spot pressure with changing thermal conductivity is found to be approximately true \cite{betti2001}, but requires a more thorough discussion. The hot-spot pressure is equivalent to the hot-spot energy density; while thermal conduction is transfer of heat from the hot-spot to the cold fuel, that energy is then recycled into the hot-spot by ablation. Therefore, varying the thermal conductivity does not change the energy density. A second-order effect is the work done by that ablating plasma on the hot-spot \cite{betti2001}; indeed, an applied magnetic field is found in simulations to marginally reduce the hot-spot pressure by reducing the work done by ablating plasma. More importantly, an applied magnetic field results in an asymmetric stagnated hot-spot \cite{perkins2017,walsh2019}; the impact of this inherent mode 2 asymmetry on hot-spot pressure will be investigated in section \ref{sec:shape}. For now, the hot-spot pressure is assumed to be unchanged by magnetization of the electrons.
	
	With these assumptions, the scaling of density and temperature amplification due to changes in thermal conductivity are found:
	
	\begin{equation}
		\frac{\rho_B}{\rho_{B=0}} = \kappa_{eff}^{2/7} \label{eq:rhoamp}
	\end{equation}
	
	\begin{equation}
		\frac{T_B}{T_{B=0}} =  \kappa_{eff}^{-2/7} \label{eq:Tamp}
	\end{equation}
	
	i.e. when thermal conduction losses are suppressed ($\kappa_{eff}$ is decreased), there is less ablation of plasma into the hot-spot (resulting in a lower density), but the reduced heat-flow thermally insulates the hot-spot to increase the temperature. This scaling has previously been used to assess the sensitivity of hot-spot temperature to thermal conductivity \cite{2018}; that scaling used the SESAME thermal conductivities \cite{lyon1995} (resulting in different temperature and density exponents) while here the form given by Spitzer-Harm is assumed \cite{spitzer1953}. The differences between thermal conductivity models will be discussed at the end of this section.
	
	The scaling of temperature with effective thermal conductivity is first done without any consideration of a magnetic field. 1-D simulations of cylindrical and spherical implosions are conducted, modifying the electron thermal conductivity in the fuel after the first shock has converged on axis (marking the beginning of the stagnation phase). In this way, $\kappa_{eff}$ is chosen for each simulation, and can have a value below 1/3 in spherical implosions, as there is no consideration of magnetic field geometry.
	
	Figure \ref{fig:Ti_kappamod} shows the temperature amplification by reducing the effective thermal conductivity from its nominal value of 1 to a series of values between 0,1. Other sources of fuel ablation (such as radiation transport and ion thermal conduction) have been included in these simulations, but $\alpha$-heating has been neglected. All designs and geometries roughly follow the $\kappa_{eff}^{-2/7}$ scaling, with greatest deviations for low values of $\kappa_{eff}$. 
	
	As $\kappa_{eff}$ tends towards zero, the simple model breaks down. The initial hot-spot mass $M_0$ was assumed to be much less than the final hot-spot mass $M_{hs}$; with $\kappa_{eff}=0$ there is theoretically no ablation of mass into the hot-spot, giving no final hot-spot mass. Likewise, the temperature given by the scaling rises to infinity. In reality the initial hot-spot mass is not negligible and other sources of energy flux (in particular radiation transport) also contribute to mass ablation into the hot-spot, even when $\kappa_{eff}=0$.
	
	The hot-spot pressure is found to be relatively unaffected by the decrease in hot-spot ablation. For the warm symcap design, the pressure decreases by 0.02\% for $\kappa_{eff}=1/3$ and a total of 0.1\% for $\kappa_{eff}=0$. For the layered design, these changes are even smaller. The more important impact on the hot-spot pressure is from the induced shape asymmetry, which will be discussed in section \ref{sec:shape}.
	
	Next, full 2-D extended-MHD simulations have been executed for a variety of applied field strengths and target designs. No perturbations have been added to these simulations apart from the natural asymmetry arising from the application of a magnetic field \cite{perkins2017,walsh2019}. Now, instead of prescribing a specific $\kappa_{eff}$, the $\kappa_{eff}$ must be calculated from the simulations. As the hot-spot elongates along the magnetic field lines \cite{walsh2019} there is an increased surface area that is perpendicular to the magnetic field. This means that a magnetized spherical capsule can reach $\kappa_{eff}<1/3$, and the hot-spot is more insulated than if it remained round. In the simulations, the following definition for $\kappa_{eff}$ is used:
	
	\begin{equation}
		\kappa_{eff} = \frac{A_z}{A_{total}} + \Big(1-\frac{A_z}{A_{total}} \Big) \frac{\kappa_{\bot}}{\kappa_{\parallel}} \label{eq:kappa_eff}
	\end{equation}
	
	Where $A_{total}$ is the total hot-spot surface area at bang-time, using the $T_e = 1keV$ contour. $A_z = \int_S |\underline{\hat{n}} \cdot \underline{\hat{z}}| dS$ is the surface area along the direction of applied magnetic field, where $S$ is the hot-spot surface and $\underline{\hat{n}}$ is the unit normal. For a completely spherical implosion, $A_z/A_{total} = 1/3$.  $\kappa_{\bot}/\kappa_{\parallel}$ is calculated using the burn-averaged electron magnetization. 
	
	Figure \ref{fig:Ti_xMHD} plots ion temperature amplification against the post-processed effective thermal conductivity (equation \ref{eq:kappa_eff}) for the spherical implosion designs. Applied fields of 0T, 5T, 10T, 20T, 30T, 40T and 50T have been used. Again, the points roughly follow the theoretical scaling of $T_i \sim \kappa_{eff}^{-2/7}$. 
	
	Some of the indirect-drive simulations in figure \ref{fig:Ti_xMHD} have $\kappa_{eff}<1/3$; this is due to the hot-spot elongation along the magnetic field lines increasing the surface area that is perpendicular to the magnetic field ($A_z/A_{total} < 1/3$). For a 50T applied field, the layered design has  $A_z/A_{total} = 0.294$ and the symcap design has $A_z/A_{total} = 0.292$ at bang-time. The symcap is also more magnetized, with burn-averaged $\kappa_{\bot}/\kappa_{\parallel} = 0.022$ as opposed to the layered target with $\kappa_{\bot}/\kappa_{\parallel} = 0.033$. Both of these contribute to the symcap temperature amplification due to a 50T magnetic field being higher than the layered target, with values of $T_B/T_{B=0} = 1.441$ and $T_B/T_{B=0} = 1.407$ respectively.
	
	The hot-spot magnetization also reduces radiative emission. The Bremsstrahlung losses scale as $P_{hs}^2/T_{hs}^{3/2}$. As the magnetic field increases the temperature but leaves the pressure approximately the same, the magnetized capsules have reduced losses, which enhances the temperature further. An equivalent set of simulations as those in figure \ref{fig:Ti_xMHD} have been executed without any radiative losses; these simulations consistently sit below those with the radiation included. 
	
	The effective thermal conductivity in equation \ref{eq:kappa_eff} can be approximated using experiment observables. The ratio $A_z/A_{total}$ can be inferred using neutron or x-ray emission images \cite{merrill2012,mcglinchey2018}. The compressed magnetic field and hot-spot density can be approximated by using the convergence ratio of the implosion. Finally, the emitted neutron spectra can be used to infer the hot-spot temperature. The combination of density, temperature and magnetic field strength then give the hot-spot magnetization.
	
	No cylindrical simulations have been included in figure \ref{fig:Ti_xMHD}. This is because the design simulated has a significant magnetic pressure in the hot-spot \cite{10.1088/1361-6587/ac3f25}, which results in lowered stagnated thermal pressures. The resultant temperature amplification falls far below the $\kappa_{eff}^{-2/7}$ scaling.
	
	Equations \ref{eq:rhoamp} and \ref{eq:Tamp} assumed a Spitzer thermal conductivity \cite{spitzer1953}. If, instead, a SESAME thermal conductivity is utilized \cite{lyon1995} the temperature scales as $\frac{T_B}{T_{B=0}} =  \kappa_{eff}^{-1/3}$ \cite{2018}. For $\kappa_{eff}=1/3$ this gives a temperature amplification of 44\% rather than 37\%.

	
	\section{Yield Amplification \label{sec:Y_amp}}
	
	\begin{figure}
			\centering
			\includegraphics[width=0.5\textwidth]{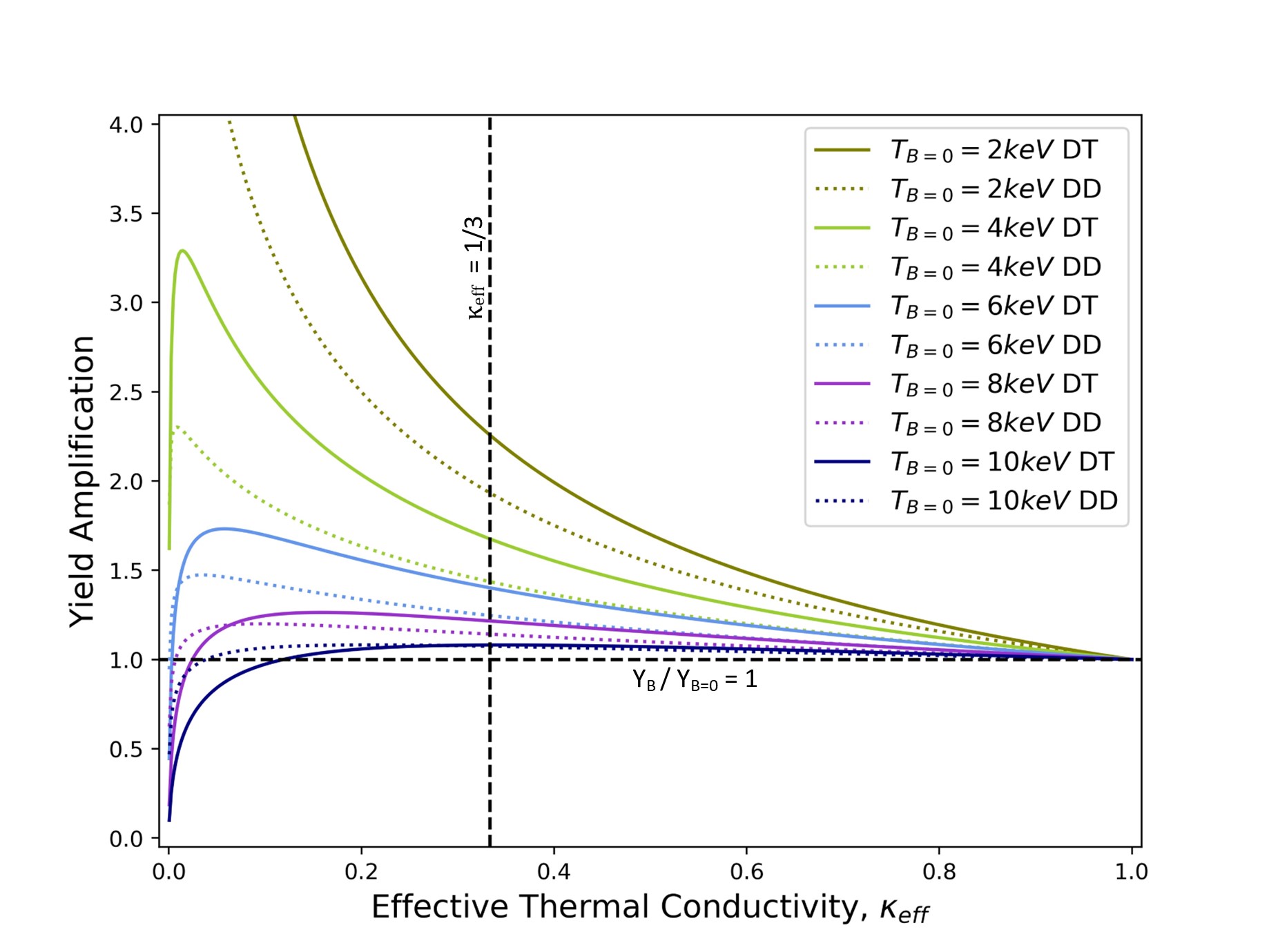}
			\caption{ Theorized increase in hot-spot yield due to the reduction of electron thermal conductivity in the fuel by factor $\kappa_{eff}$. Results for both DD and DT reactions are shown, using the Bosch-Hale formulation\cite{1993}.\label{fig:Yamp_theory} }
		\end{figure}
		
		\begin{figure}
			\centering
			\includegraphics[width=0.5\textwidth]{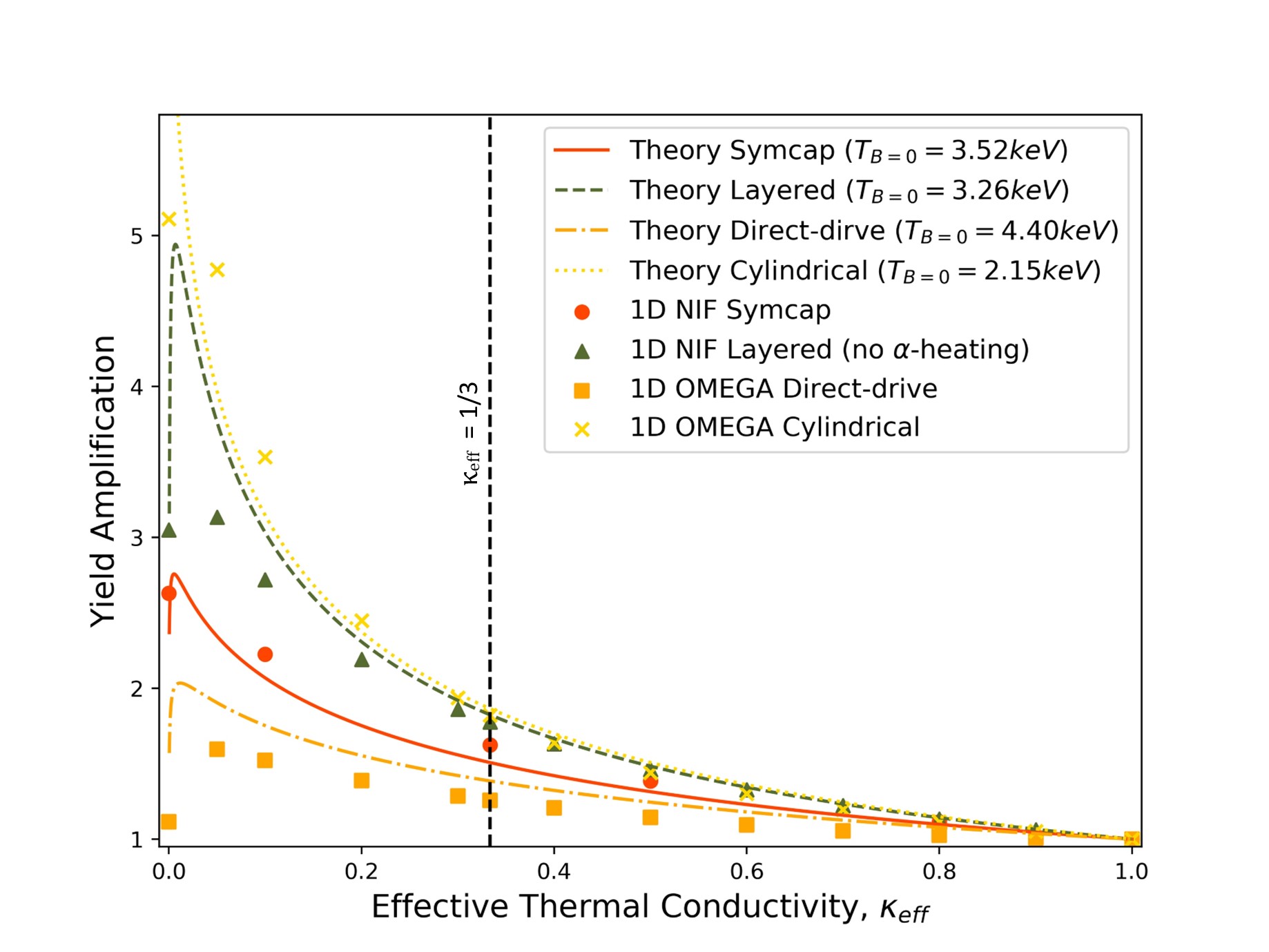}
			\caption{ \label{fig:Yamp_kappamod} Yield amplification by reducing the fuel electron thermal conductivity by a factor $\kappa_{eff}$. Both theory (lines) and 1-D simulations (points) are shown. }
	\end{figure}
	
	Using the scaling of temperature amplification by magnetization outlined in section \ref{sec:temp_amp}, the theorized amplification of fusion reactions can be explored. In this section the discussion is limited to unperturbed capsules, while sections \ref{sec:shape}, \ref{sec:P4_shape} and \ref{sec:pert} investigate P2 shape, P4 shape and higher mode perturbations.
	
	The number of reactions between species $a$ and $b$ over the burn time-scale $\delta t_{burn}$ can be approximated as:
	
	\begin{equation}
		Y_{ab} \approx \delta t_{burn} \frac{4 \pi R_{hs}^3}{3} <\sigma_{ab} v > \frac{P_i^2}{(2-\delta_{ab})^2T_i^2} \label{eq:Yab}
	\end{equation}

	Where $<\sigma_{ab} v >$ is the reactivity between species $a$ and $b$. $\delta_{ab}$ is the Kronecker delta, which takes the values $\delta_{ab}=1$ for $a=b$ and $\delta_{ab}=0$ for $a\ne b$; this represents the higher availability of ions for single-species reactions.
	
	Similar to section \ref{sec:temp_amp}, it is assumed that the magnetic field does not affect the burn time-scale, hot-spot radius or hot-spot pressure. Section \ref{sec:shape} explores the dependence of hot-spot pressure on magnetization in more detail; as magnetized hot-spots tend to elongate along magnetic field lines \cite{walsh2019}, a P2 drive asymmetry is needed to regain much of the hot-spot pressure lost by magnetization.
	
	By combining the scaling of yield (equation \ref{eq:Yab}) with a simplified form for the fusion reactivity $<\sigma_{ab} v >$ \cite{huba2013} and the scaling of temperature with effective thermal conductivity (equation \ref{eq:Tamp}), a yield amplification for DT and DD reactions by electron magnetization can be found:
	
	\begin{equation}
		\Bigg[\frac{Y_B}{Y_{B=0}}\Bigg]_{DT} =  \kappa_{eff}^{16/21} \exp \Bigg( 19.94 \frac{1-\kappa_{eff}^{2/21}}{T_{B=0}^{1/3}} \Bigg) \label{eq:Yamp_DT}
	\end{equation}
	
	\begin{equation}
		\Bigg[\frac{Y_B}{Y_{B=0}}\Bigg]_{DD} =  \kappa_{eff}^{16/21} \exp \Bigg( 18.76 \frac{1-\kappa_{eff}^{2/21}}{T_{B=0}^{1/3}} \Bigg) \label{eq:Yamp_DD}
	\end{equation}

	While temperature amplification by magnetization is dependent only on the electron Hall Parameter (through $\kappa_{eff}$), the yield amplification is also dependent on the unmagnetized hot-spot temperature ($T_{B=0}$). Equations \ref{eq:Yamp_DT} and \ref{eq:Yamp_DD} can be improved by using the Bosch-Hale formulation \cite{1993}, rather than the simplified form; Bosch-Hale is used for all theory and simulations in this paper, but the resultant yield amplification equations are unwieldy (albeit simple to derive). Equations \ref{eq:Yamp_DT} and \ref{eq:Yamp_DD} demonstrate all the important Physics and give approximately the same answer as Bosch-Hale, albeit under-predicting the benefits of magnetization by up to 20\% in regimes of interest. 
	
	Figure \ref{fig:Yamp_theory} plots the theorized yield amplification as a function of effective thermal conductivity for a range of unmagnetized hot-spot temperatures. It is worth re-iterating that the theory assumes the same hot-spot pressure for both magnetized and unmagnetized cases; as the temperature increases, the density decreases. For the yield to go up with magnetization, the decrease in yield due to lower ion densities must be overcome by the increase in fusion reactivity with temperature. The DT reactivity increases quicker with temperature than the DD, making magnetization more beneficial for DT experiments in the regime of interest. 
	
	Hot-spots with initially low temperatures benefit most strongly from magnetization. For $\kappa_{eff}=1/3$, which is appropriate for a highly magnetized spherical implosion, a DT hot-spot with unmagnetized temperature $T_{B=0} = 2$keV has an expected yield amplification $Y_{B}/Y_{B=0} = 2.3$, while a hot-spot of unmagnetized temperature $T_{B=0} = 6$keV would expect a yield amplification of $Y_{B}/Y_{B=0} = 1.4$. For a high performing hot-spot with $T_{B=0} = 10$keV the anticipated 37\% increase in temperature by magnetization has very little effect on the yield, with $Y_{B}/Y_{B=0} = 1.08$. Note that none of these results include any consideration of $\alpha$-heating, which tends to amplify a yield enhancement \cite{Tong_2019}. 
	
	Figure \ref{fig:Yamp_theory} shows it is possible to decrease the yield by magnetization. By looking at a simple formula for fusion reactions it is possible to show that the highest yield is reached when the hot-spot temperature is increased to 15.5keV for DT reactions (12.9keV for DD) \cite{huba2013}. Above this temperature the decrease in ion density by magnetization degrades the yield faster than the increased reactivity.
	
	The yield scaling is tested against 1-D simulations where the thermal conductivity in the fuel is varied by a chosen factor $\kappa_{eff}$. These calculations do not include any MHD physics, but allow for a clean comparison with the scaling without shape effects and magnetic transport complicating matters. 
	
	Figure \ref{fig:Yamp_kappamod} shows the yield amplification ($Y_{\kappa_{eff}}/Y_{\kappa_{eff}=0}$) for NIF indirect-drive designs (layered\cite{clark2019} and a symcap\cite{moody2021,moody2021a,casey2018}), a layered OMEGA direct-drive implosion\cite{PhysRevLett.100.185006} and an OMEGA cylindrical configuration\cite{hansen2020,10.1088/1361-6587/ac3f25}; the simulations are the same as those used in figure \ref{fig:Ti_kappamod}. For each design a different theoretical scaling of yield amplification is plotted, as each design has a different unmagnetized hot-spot temperature. Both indirect-drive and direct-drive layered designs use DT fuel while the NIF symcap and the cylindrical implosions use DD.
	
	The cylindrical design is found to improve most from reduced thermal conductivities, as the initial hot-spot temperature is only 2.15keV. At this low temperature the DD reactivity increases rapidly with temperature.
	
	The yield scalings compare favorably to the 1-D simulations. Note that the theory breaks down for $\kappa_{eff} = 0$, predicting that the hot-spot temperature increases to infinity.
	
	While the theoretical yield enhancement works effectively for 1-D simulations,  2-D MHD calculations differ greatly from these simple expectations. Section \ref{sec:shape} shows that the hot-spot shape is the primary reason for this discrepancy; hot-spot elongation along magnetic field lines results in a lower hot-spot pressure. It is shown that an asymmetric drive can be used to correct for this natural asymmetry with an applied magnetic field. 
	

	\section{P2 Shape \label{sec:shape}}
	
		\begin{figure}
		\centering
		\begin{subfigure}[b]{0.5\textwidth}
			\centering
			\includegraphics[width=1.\textwidth]{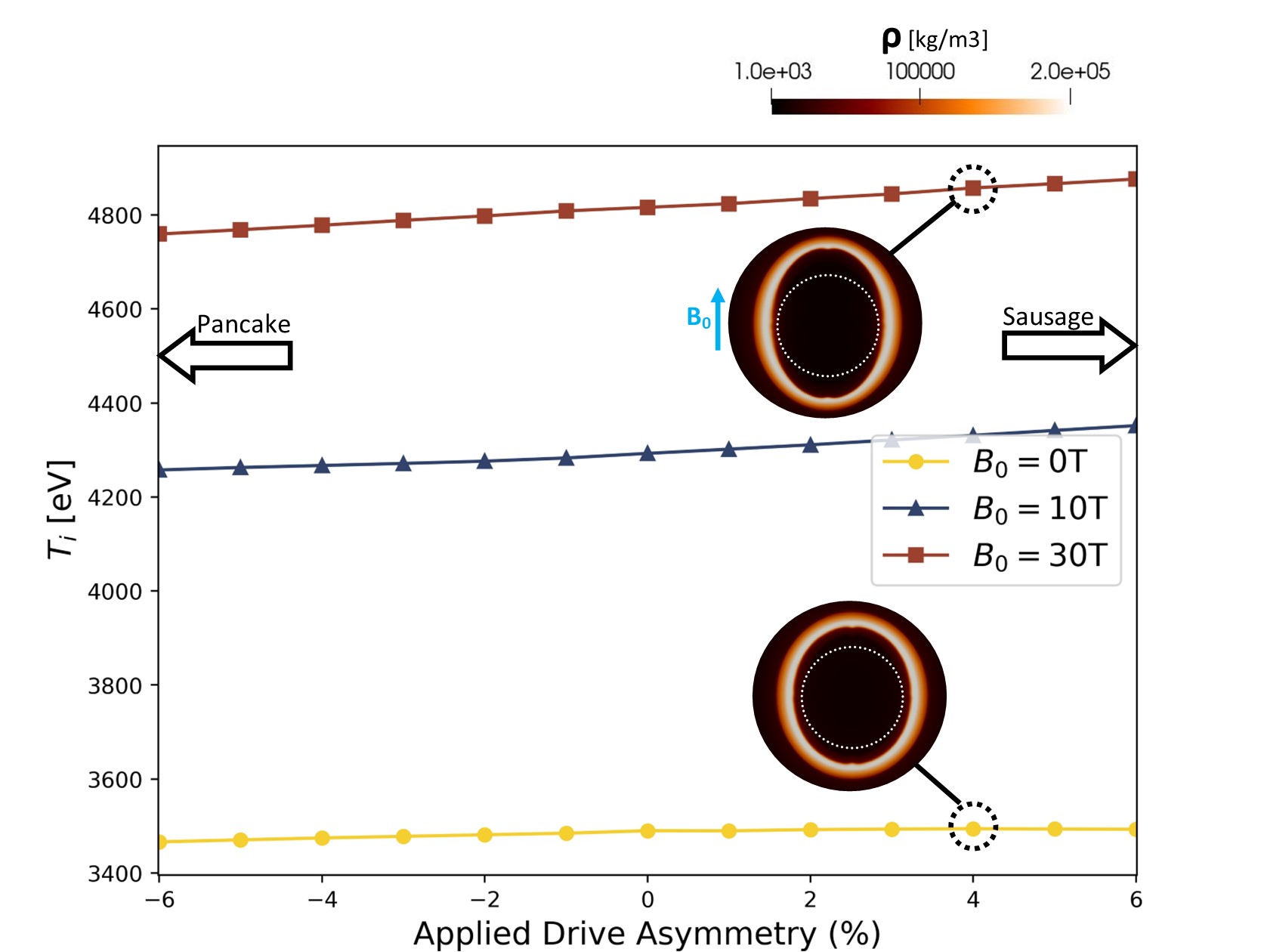}
			\caption{}
			\label{fig:P2_Ti}
		\end{subfigure}
		\begin{subfigure}[b]{0.5\textwidth}
			\centering
			\includegraphics[width=1.\textwidth]{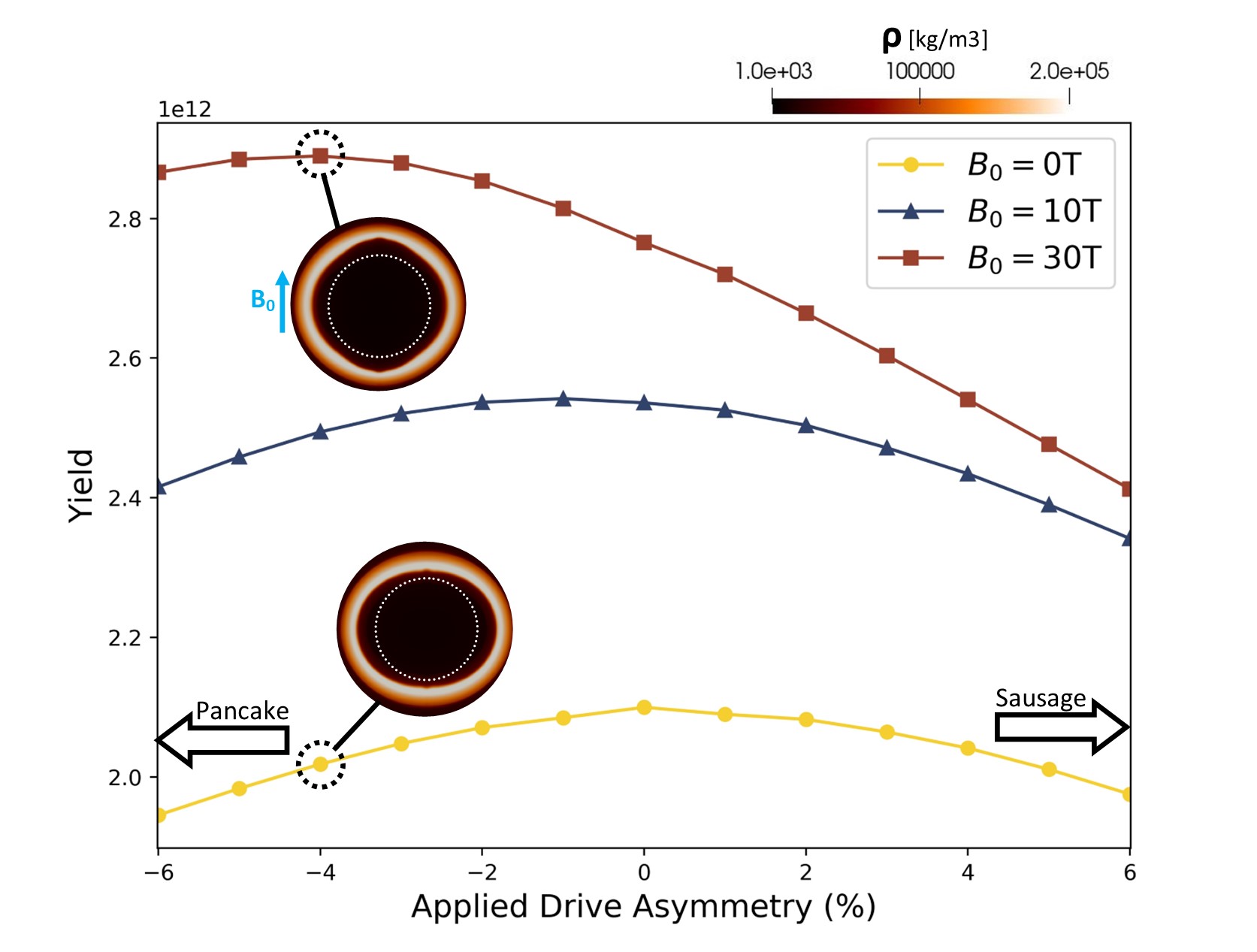}
			\caption{}
			\label{fig:P2_Y}
		\end{subfigure}
		\begin{subfigure}[b]{0.5\textwidth}
			\centering
			\includegraphics[width=1.\textwidth]{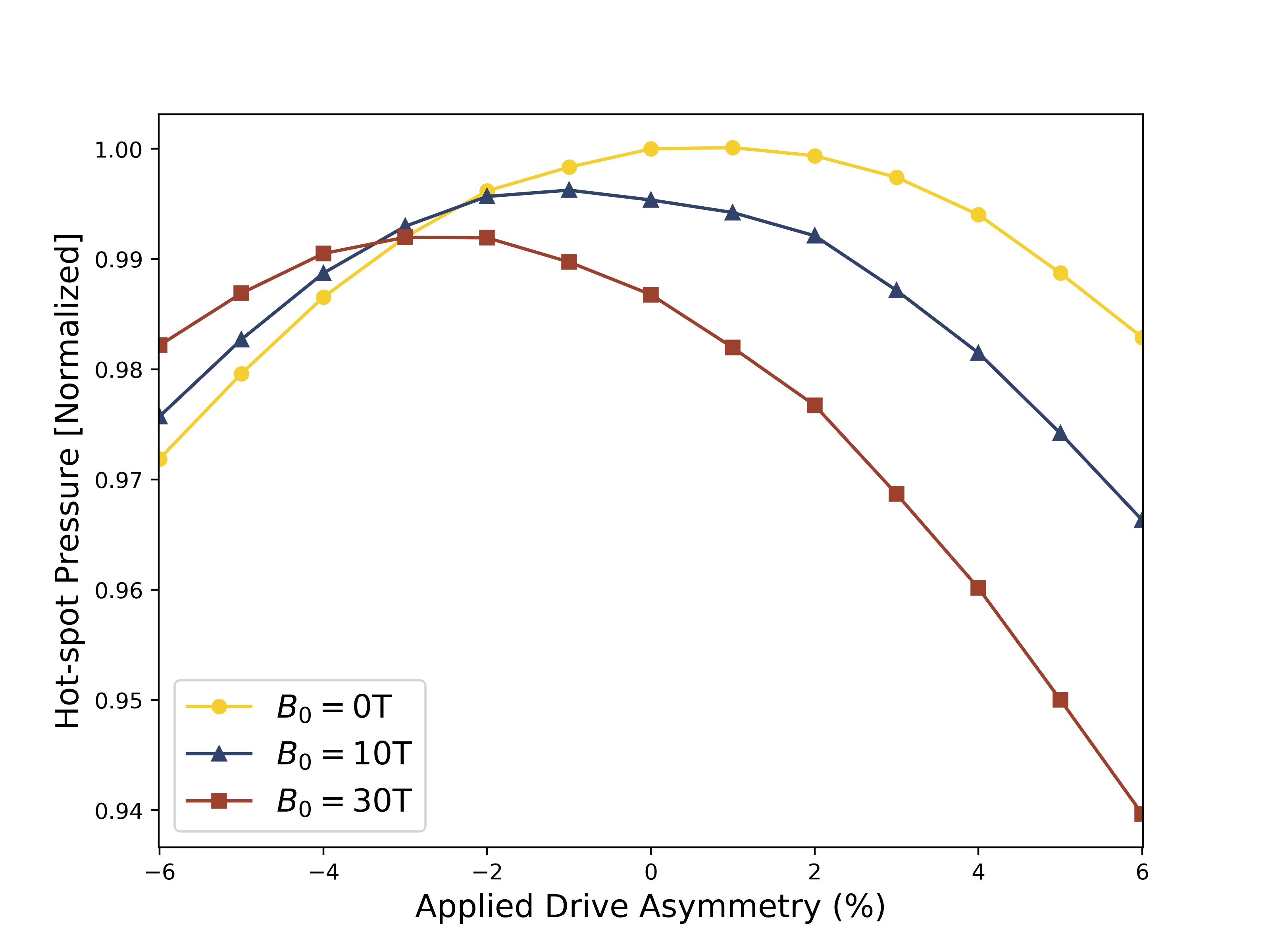}
			\caption{The dependence of burn-averaged hot-spot temperature, DD yield and pressure on the applied P2 drive asymmetry for 0T, 10T and 30T initial field strengths. Demonstrative density slices at bang-time are overlain, with white dashed circles added to assist in assessing the final hot-spot symmetry. The 0T case is most symmetric for no drive asymmetry, while the 30T case is closest to round for a +4\% P2 drive.}
			\label{fig:P2_P}
		\end{subfigure}
		\caption{}
		\label{fig:P2}
	\end{figure}

	This section investigates mode 2 asymmetries that share the same axis as the applied magnetic field, i.e. a Legendre P2 mode. As an applied magnetic field does not suppress heat-flow along the field direction, a P2 naturally forms, even when the capsule is symmetrically driven \cite{walsh2019}. How the unmagnetized and magnetized performance vary with a purposefully applied P2 drive asymmetry will also be investigated. This section focuses on the warm NIF symcap design. The drive asymmetry is a constant percentage in time, which does not represent the natural changes in drive symmetry throughout a hohlraum's history; this is intended as a simple test problem. For indirect-drive implosions the P2 shape is routinely controlled through laser cone fraction or cross-beam energy transfer \cite{PhysRevLett.102.025004,PhysRevLett.111.235001}. On OMEGA low-mode asymmetries have also been demonstrated \cite{PhysRevE.98.051201,gatujohnson2019}.

	Figure \ref{fig:P2_Ti} shows the ion temperature as a function of drive asymmetry for 0T, 10T and 30T applied magnetic fields. Density plots at bang-time are shown to give a reference point for the level of asymmetry in the implosions; it can be seen that the 30T field increases the elongation of the hot-spot. White dashed circles have been included to aid the eye in seeing the level of asymmetry. 
	
	The magnetized cases increase in temperature when the P2 drive is increased (making them more sausage). This is consistent with the findings in section \ref{sec:temp_amp}, where it was posited that the hot-spot elongation reduces the effective thermal conductivity by giving a larger proportion of the hot-spot area perpendicular to the magnetic field lines. For this case the change in temperature with P2 drive is small, increasing by 2.6\% when going from -6\% to +6\% drive P2.
	
	Increasing the hot-spot temperature does not universally increase the yield. Figure \ref{fig:P2_Y} shows how the yield varies with drive asymmetry. The result is opposite to the temperature dependence, with the capsule yield peaking when the magnetized implosion is driven with a negative P2 (to make the implosion more pancake shaped). For all levels of magnetization the implosion yield peaks when the hot-spot is the most round at bang-time. Figure \ref{fig:P2_P} shows how the hot-spot pressure (taken as a burn-averaged value) varies with the applied drive. For a symmetric drive the 30T pressure is substantially below the unmagnetized case. Driving the 30T case to be round mostly recuperates the losses of having an asymmetric hot-spot. Note that only the P2 component has been tuned out in this implosion; the 30T density plot in figure \ref{fig:P2_Y} also contains a P4 component which could be accounted for. 
	
	
	
	Once the 30T hot-spot is corrected to be round (using a -4\% P2 drive asymmetry in this case), the agreement with the theoretical yield amplification improves. The effective thermal conductivity from the simulation is calculated as 0.386, giving a predicted yield amplification of 1.44. Using the round unmagnetized implosion as the base case, the yield amplification from the 2-D simulations is 1.38. The 30T simulation without correcting for shape gives a 1.31 yield amplification.
	
	Therefore, it is asserted that the theoretical yield amplification due to magnetization (equations \ref{eq:Yamp_DD} and \ref{eq:Yamp_DT}) is approximately true when both the magnetized and unmagnetized hot-spots are round (each requiring a different drive asymmetry).

	\section{P4 Shape \label{sec:P4_shape}}
	
	Axi-symmetric mode 4 asymmetries (P4 Legendre modes) are commonly observed in indirect-drive implosion on the National Ignition Facility \cite{clark2016,mcglinchey2018}. The axi-symmetry of these perturbations allows them to be simulated in 2-D even when a magnetic field is applied \cite{walsh2021magnetized}. By varying the P4 drive amplitude, the impact of asymmetries on the magnetized temperature and yield amplification can be explored. While section \ref{sec:pert} also studies this behavior for higher modes, those modes are not axi-symmetric in experiments and therefore should be modeled in 3-D \cite{walsh2021magnetized}.
	
	For this section the layered NIF DT design is utilized \cite{clark2019}, as the high convergence results in greater importance of the magnetic tension for perturbation growth \cite{walsh2019}.
	
\begin{figure}
		\centering
		\begin{subfigure}[b]{0.5\textwidth}
			\centering
			\includegraphics[width=1.\textwidth]{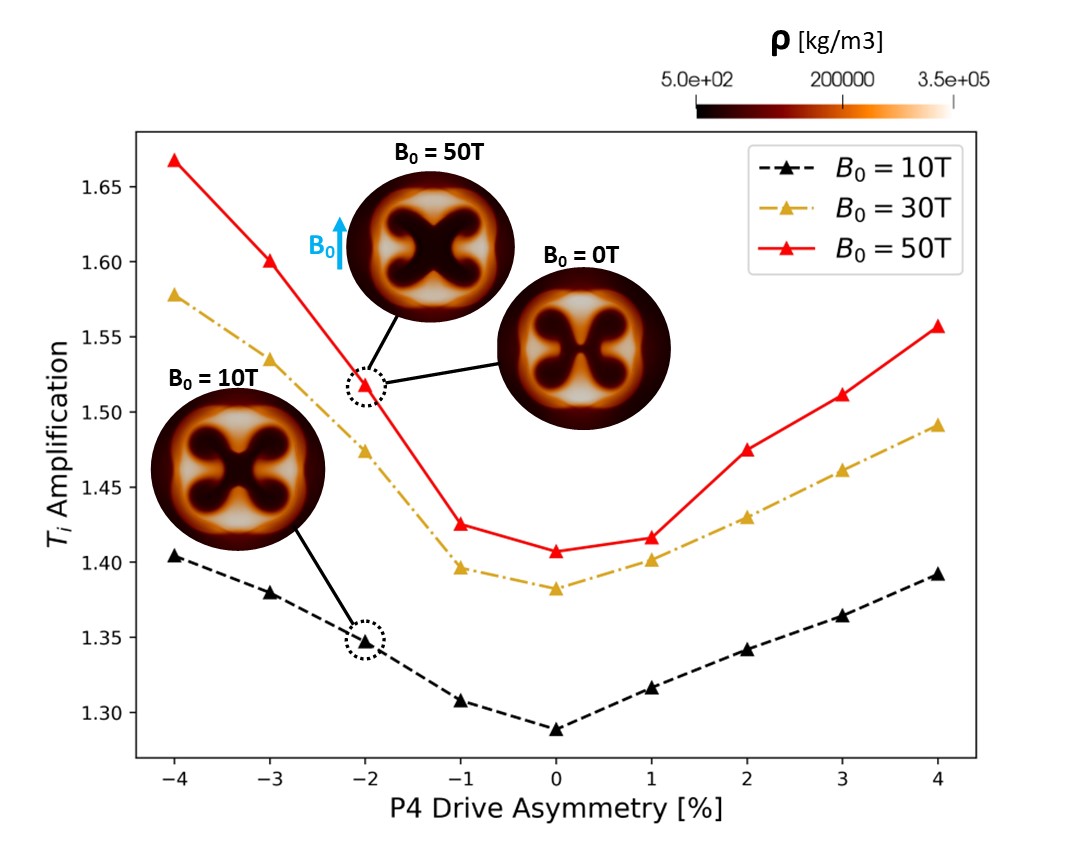}
			\caption{}
			\label{fig:P4_Ti}
		\end{subfigure}
		\begin{subfigure}[b]{0.5\textwidth}
			\centering
			\includegraphics[width=1.\textwidth]{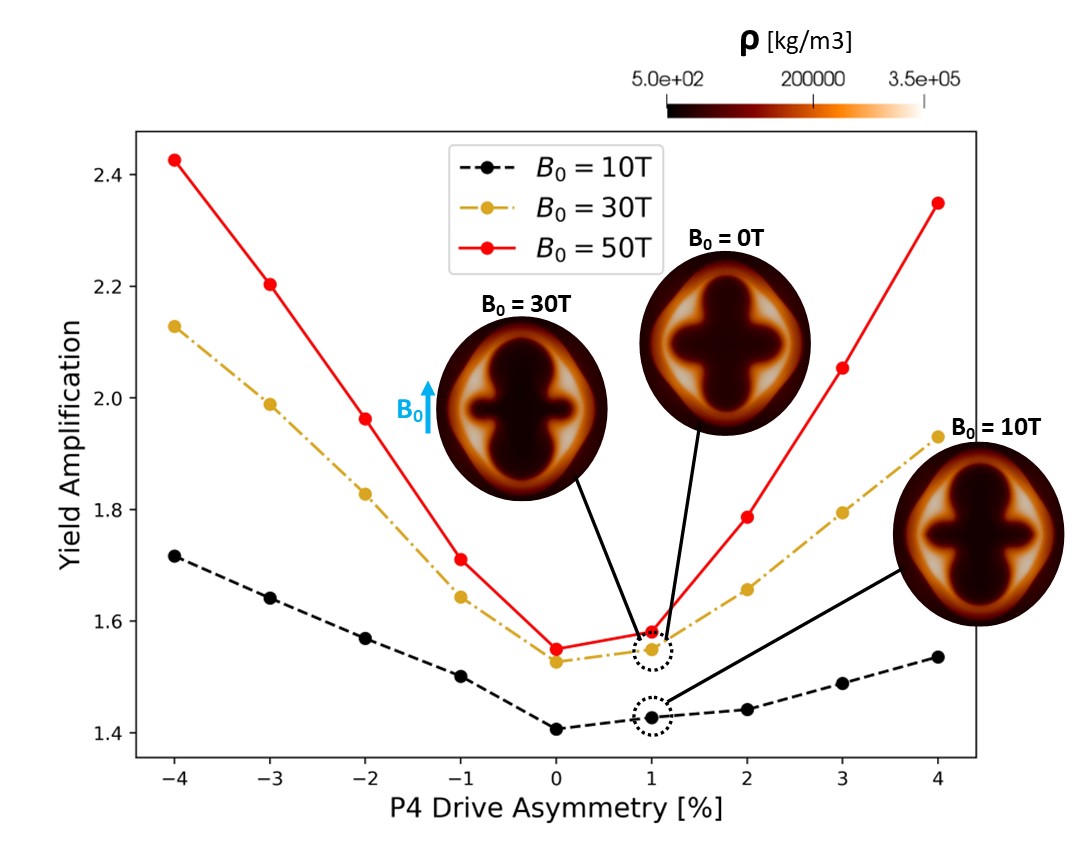}
			\caption{}
			\label{fig:P4_Y}
		\end{subfigure}
		\caption{Temperature and yield amplification from an applied magnetic field for layered NIF simulations with a range of P4 drive asymmetries. Demonstrative density slices at neutron bang-time are overlaid. As the capsules are more perturbed, the benefit of the magnetic field increases.}
		\label{fig:P4}
	\end{figure}

	Figure \ref{fig:P4} shows the temperature and yield amplification due to magnetization as a function of P4 drive asymmetry for 10T, 30T and 50T applied magnetic fields. Example density plots at bang-time are also shown for specific drive asymmetries. A negative P4 drive asymmetry results in spikes propagating into the hot-spot at the poles and cold plasma constricting around the hot-spot at the waist. Positive P4 results in cold plasma constricting at $\theta = \pi/4,3\pi/4$.
	
	In all cases a larger asymmetry results in lower temperature and yield. For a symmetric drive the unmagnetized yield is 6.1$\times$10$^{15}$, decreasing by 63\% for a -4\% P4 drive and 68\% for +4\% P4 drive. The magnetized cases also decrease in temperature and yield as the drive is more asymmetric.

	The benefit of the magnetic field, however, is found to be greatest for more perturbed capsules. The theorized yield amplification $Y_{B}/Y_{B=0}$ (from figure \ref{fig:Yamp_kappamod}) for the 50T symmetric drive case is 1.84. As discussed in section \ref{sec:shape}, the simulated yield amplification falls far below the theory ($Y_{B}/Y_{B=0}=$1.56) due to hot-spot elongation lowering the pressure. Once the capsule is  perturbed with a P4, the yield enhancement can greatly exceed the symmetric theory, with  $Y_{B}/Y_{B=0}=$2.43 for a -4\% P4 drive.
	
	Yield and temperature amplification are more greatly enhanced by magnetization for negative P4 asymmetries. This is for 2 reasons. Firstly, the cold plasma constricting around the waist is stabilized by magnetic tension effects, meaning that the hot-spot core remains more open when a larger magnetic field is applied \cite{walsh2019}. Secondly, the enhanced hot-spot temperature by magnetization results in larger heat-flows along the magnetic field lines to the capsule poles; this results in greater thermal ablative stabilization of the spikes that propagate down the poles \cite{walsh2019}. Both of these effects can be seen by comparing the 0T and 50T density slices in figure \ref{fig:P4_Ti}.
	
	Increasing the applied field strength from 30T to 50T is also more beneficial once the capsule is perturbed. For a symmetric drive the yield amplification goes from 1.41 to 1.54 when 30T is used rather than 10T. Increasing the field further has little impact, with $Y_{B}/Y_{B=0}=$1.56 for 50T. This is because the 30T hot-spot is already very magnetized, with a burn averaged Hall parameter of 17, corresponding to a perpendicular thermal conductivity suppression of $\kappa_{\bot}/\kappa_{\parallel}=0.004$. 
	
	Once the capsule is perturbed, however, the magnetic tension plays an important role in improving the hot-spot symmetry. While the electron magnetization effectively plateaus once 30T is applied, the magnetic tension scales with $|\underline{B}|^2$. For a +4\% P4 drive asymmetry, the 50T field has a yield amplification by magnetization of 2.35, as compared with the 30T case of 1.92.

	\section{Higher Mode Perturbations \label{sec:pert}}
	
	\begin{figure*}
		\centering
		\includegraphics[width=0.6\textwidth]{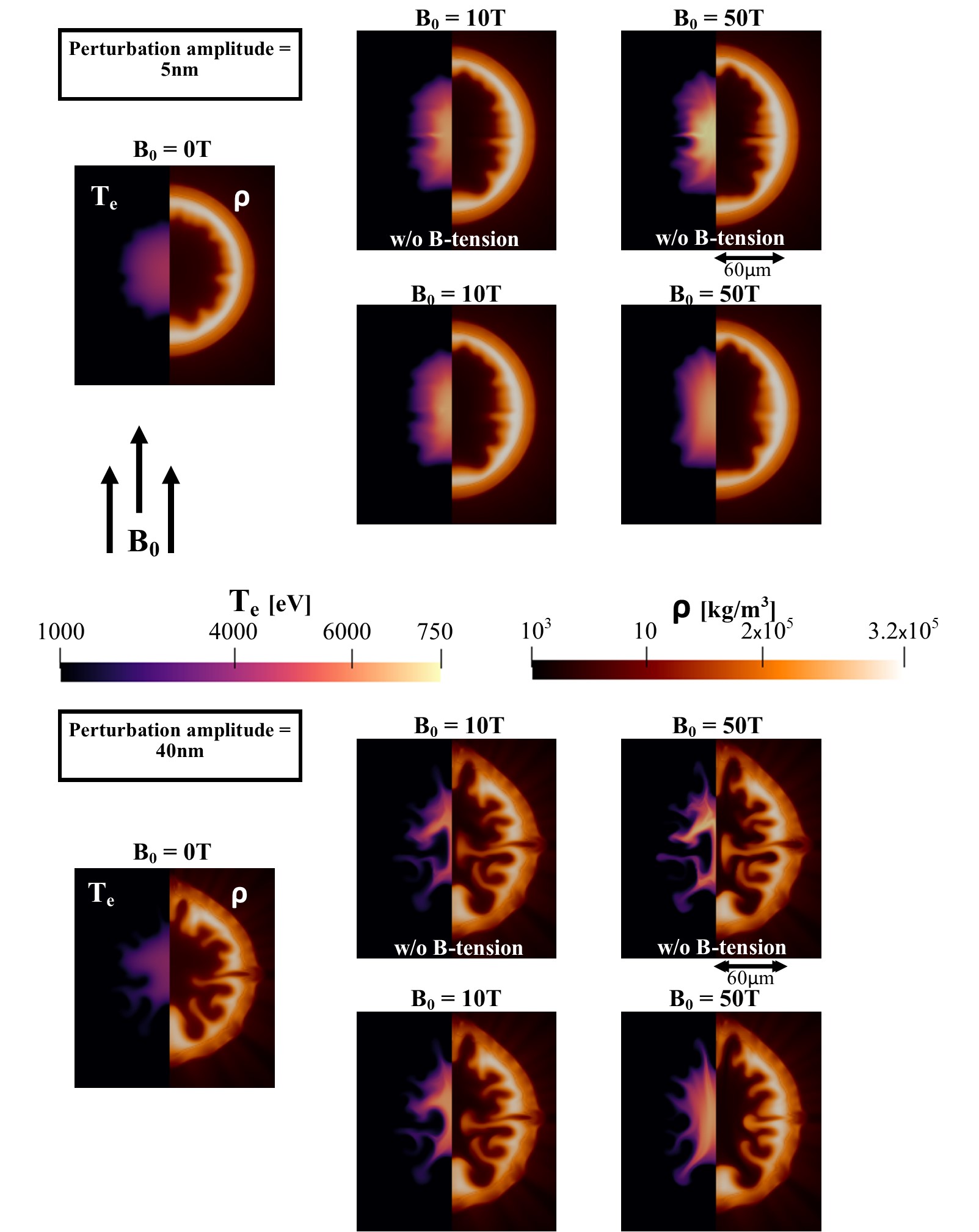}
		\caption{Electron temperature and density at neutron bang-time for 2-D simulations with multi-mode perturbations applied (top = 5nm amplitude; bottom = 40nm amplitude). Plots are shown for applied magnetic fields of 0T, 10T and 50T. For the magnetized calculations, cases both without and with the magnetic tension stabilizing force are presented, as neither case fully captures the intrinsically 3-D evolution \cite{walsh2021magnetized}. \label{fig:Pert_slices}} 
	\end{figure*}
	
	This section incorporates higher mode perturbations into the 2-D simulations to investigate how the yield and temperature amplification depend on perturbation size. The layered indirect-drive design is chosen, as the higher convergence means perturbations of this type are expected to have a significant impact on the yield \cite{clark2019}. The perturbations are initialized as HDC shell thickness asymmetries \cite{casey2021}, although the amplitudes and distributions of modes are not based on real target specifications; again, this perturbation source is used as a toy problem to investigate the general behavior of magnetized implosions. 400 modes are applied, each with amplitudes chosen randomly between 0,$a$ and with mode numbers randomly between k=1,180 with linear distribution. $a$ is varied between 0,60nm in this study. The same random numbers have been used between all the cases. This approach has also been used to investigate the self-generation of Biermann battery magnetic fields during hot-spot stagnation\cite{walsh2021a}. 
		
	2-D simulations enforcing symmetry around the magnetic field axis are used. Real perturbations with these mode numbers will develop in 3-D \cite{clark2016,clark2019}. It has previously been shown that 2-D MHD simulations of 3-D asymmetries will drastically over-predict the importance of magnetic tension stabilization \cite{walsh2021magnetized}, as the simulations do not include the dimension where perturbation growth is not stabilized, but can instead be destabilized. Sections \ref{sec:shape} and \ref{sec:P4_shape} looked at specific cases where axi-symmetry is valid; in this section, only fully 3-D calculations are valid. 3-D MHD simulations of magnetized hot-spots have been completed \cite{walsh2019}, but these calculations are too laborious to scan perturbation amplitude. In order to get a basic idea of how magnetic tension and electron magnetization scale with perturbation amplitude, simulations with both the magnetic tension turned on and off are presented. The author posits that the 3-D truth is likely bounded by these answers. 
	
	Figure \ref{fig:Pert_slices} shows density and temperature at neutron bang-time for 2 perturbation amplitudes and 0T, 10T and 50T applied fields. The magnetized simulations are shown with both magnetic tension turned on and off. For the smaller amplitude perturbations the main impact of the magnetic field is to thermally insulate the hot-spot, increasing the temperature and elongating along the magnetic field lines.
	
	For larger amplitude perturbation cases (with maximum amplitude per mode of a=40nm), the impact of magnetization on perturbation growth can be seen. From 0T to 10T the perturbations become more unstable at the waist due to suppression of thermal ablative stabilization \cite{walsh2019}. For 10T the impact of magnetic tension stabilization is small. However, the Lorentz force scales as $|\underline{B}\cdot\underline{B}|$, meaning that the 50T case has a magnetic tension stabilizing force 25 times larger than the 10T capsule. For the highly magnetized case the tension substantially suppresses the Rayleigh-Taylor growth at the waist, giving a yield amplification $Y_{B=50T}/Y_{B=0} = $ 1.94.
	
	Note that the highly perturbed capsules compress magnetic field exactly onto the axis. In reality, a 3-D picture would reduce the magnetic flux compression in the hot-spot core and increase the impact of residual kinetic energy.

\begin{figure}
		\centering
		\begin{subfigure}[b]{0.5\textwidth}
			\centering
			\includegraphics[width=1.\textwidth]{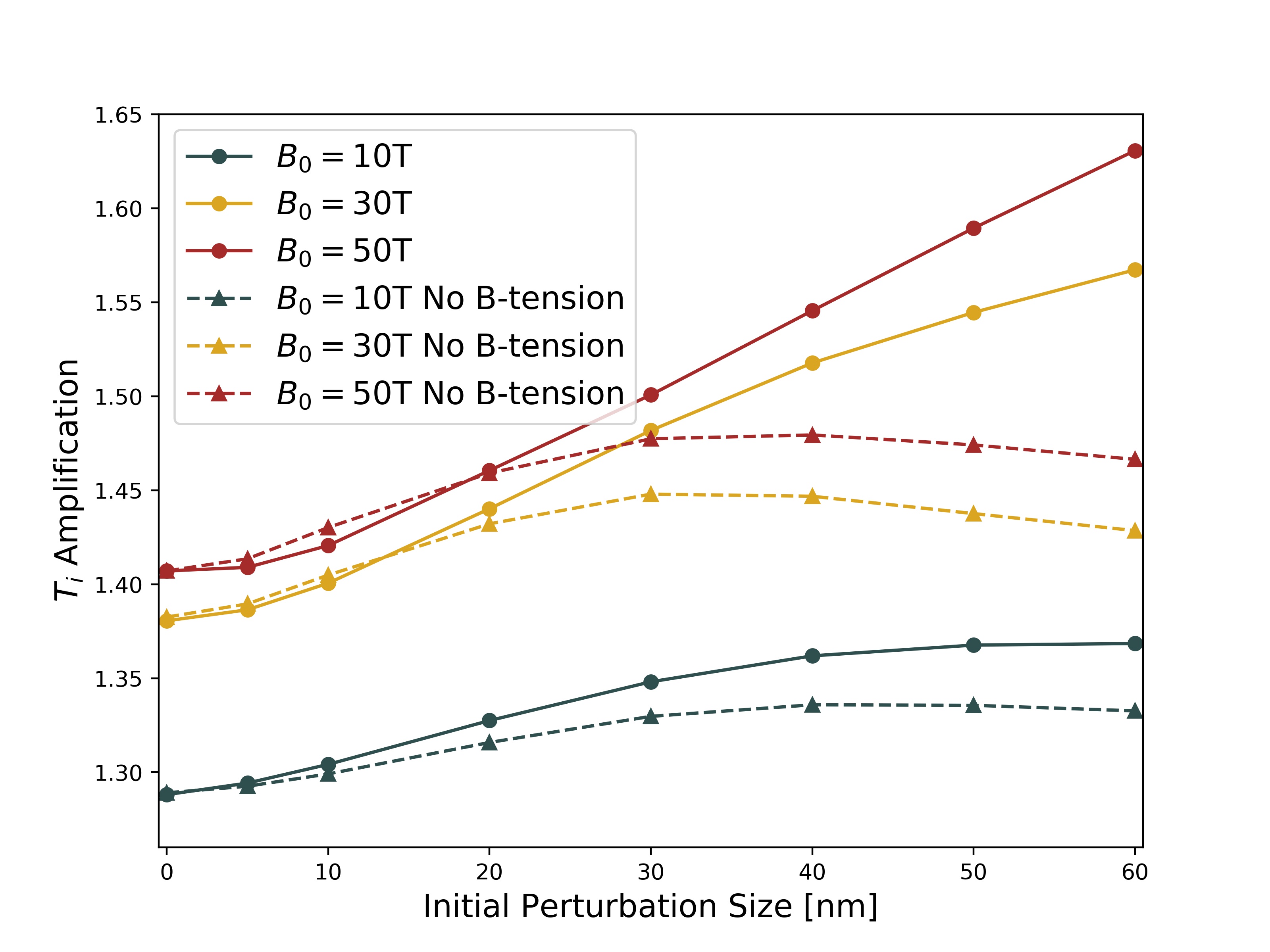}
			\caption{}
			\label{fig:Pert_Tiamp}
		\end{subfigure}
		\begin{subfigure}[b]{0.5\textwidth}
			\centering
			\includegraphics[width=1.\textwidth]{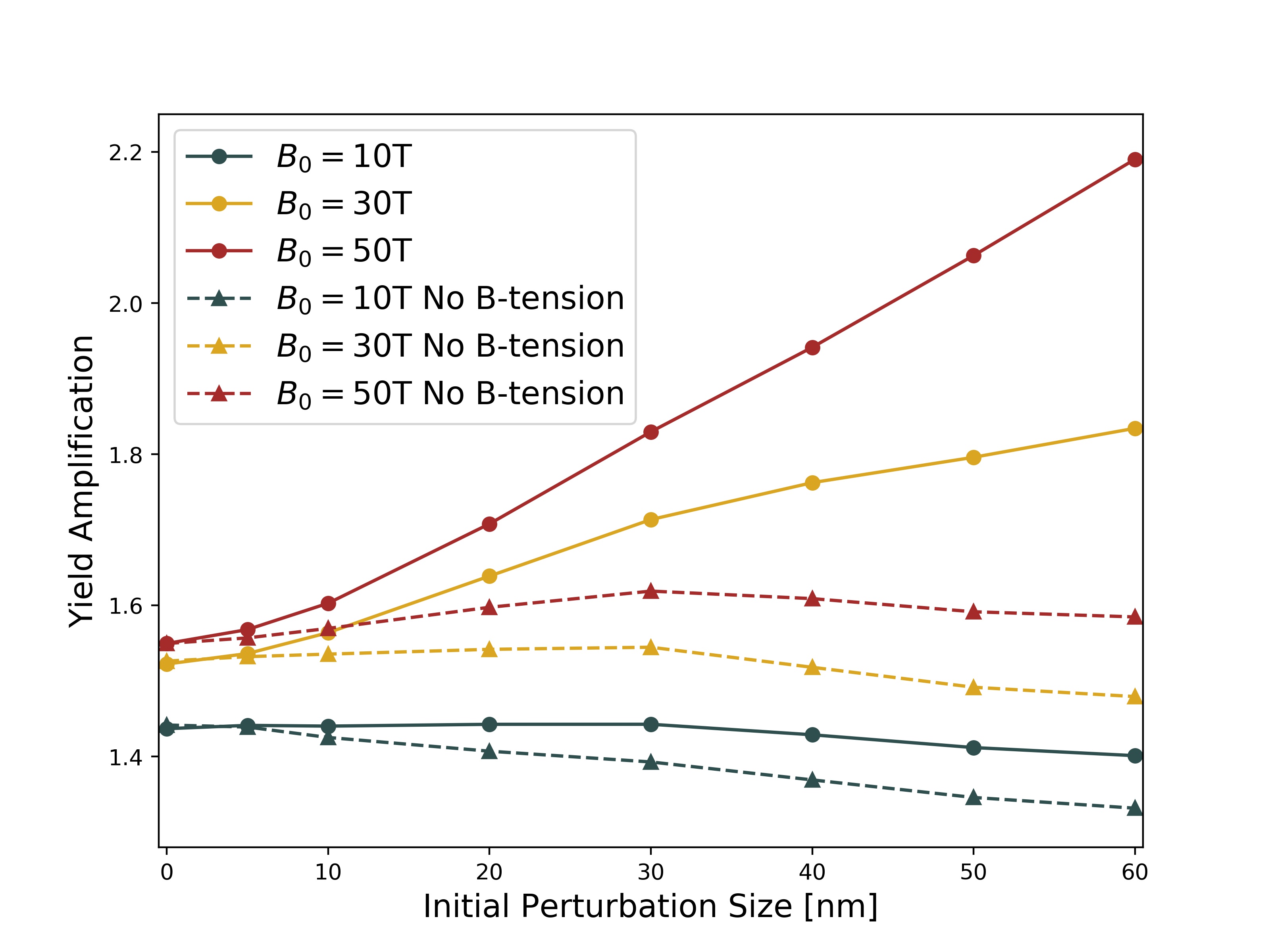}
			\caption{}
			\label{fig:Pert_Yamp}
		\end{subfigure}
		\caption{Temperature and yield amplification due to magnetization for capsules with multi-mode perturbations. Cases with and without magnetic tension stabilization are presented. The benefit of a 50T applied field over 30T is most obvious when the capsules are highly perturbed.}
		\label{fig:Pert}
	\end{figure}

	Figure \ref{fig:Pert} shows the temperature and yield amplification as a function of the initial perturbation size. The symmetric theory predicts the fully magnetized yield amplification to be $Y_{B}/Y_{B=0}=$1.84. For unperturbed implosions the yield amplification is far below the theory, with 30T and 50T giving amplifications of 1.53 and 1.55 respectively; for unperturbed capsules there is no obvious benefit for going from 30T to 50T. However, a highly perturbed implosion (a=60nm) has a simulated 50T yield amplification (with magnetic tension included) of 2.19. For 30T the yield amplification is 1.83.
	
	The impact of magnetic field on perturbation growth is highly dependent on the distributions of mode numbers in the implosion \cite{walsh2021magnetized,walsh2019}, with high modes more affected by both decreased ablative stabilization and enhanced magnetic tension stabilization. The maximum mode number applied in the presented simulations was k=180; calculations including higher mode sources (such as surface roughness) would be beneficial.

	\section{Conclusions \label{sec:conc}}
	
	A theoretical scaling of temperature amplification with hot-spot magnetization and shape has been shown to compare favorably with 2-D extended-MHD simulations of multiple magnetized implosion designs, including both directly and indirectly driven schemes. The theory assumes that the magnetic field only affects the ablation of cold fuel into the hot-spot during stagnation, lowering the hot-spot density while increasing the temperature. The theory also predicts that the temperature amplification increases when the hot-spot is more sausage shaped, which is borne out in simulations with applied P2 drive asymmetries; this is due to the smaller area with magnetic fields normal to the hot-spot surface. 
	
	Using this theory, an effective thermal conductivity of $\kappa_{eff}=0.6$ can be inferred for 8T experiments on the OMEGA Laser Facility that observed a 15\% increase in temperature \cite{chang2011}. This suggests that increasing the magnetic field strength would continue to improve the performance.  

	The theory can be used to estimate an expected yield enhancement by magnetization, with low temperature unmagnetized implosions benefiting most from an applied field. However, the theory over-predicts yield amplification due to lower hot-spot pressures when the hot-spot elongates along magnetic field lines. It is shown that a purposeful P2 drive asymmetry should be applied in order to make the implosion round again, improving the yield enhancement and bringing the results closer to the theoretical predictions.
	
	Magnetic fields have also been shown to improve the robustness of implosions to perturbations, with greater yield enhancements when the capsules are perturbed by P4 and multi-mode asymmetries. This is particularly the case when the magnetic tension plays a significant role in suppressing perturbation growth. 
	
	So far, all effort has been spent on assessing the impact of magnetic fields on current designs. Next it will be important to optimize implosions for magnetization. The results here find greatest performance enhancements for cold and highly perturbed implosions; this implies that magnetized capsules may be able to succeed at a lower adiabat than unmagnetized implosions.
	
	The results also suggest that facilities should strive for higher applied magnetic field strengths. While 30T is typically sufficient to fully magnetize the hot-spot electrons, higher fields result in the greater influence of magnetic tension, which can assist in stabilizing perturbation growth.
	
	All results in this paper are without $\alpha$-heating. If the implosion is at the foot of the ignition cliff then a burn-off yield amplification between 1.5-2 may be enough to result in much larger burn-on yield amplifications \cite{perkins2017}. An applied magnetic field can also confine $\alpha$-particles to the hot-spot, resulting in greater yield enhancements than accounted for in this paper \cite{walsh2019}. 
	
	In order to assess the impact of an applied magnetic field on the recent high yield NIF experiment N210808, a 1-D simulation with thermal conductivity artificially suppressed was completed. This is the same methodology as used for the simulations in figures \ref{fig:Ti_kappamod} and \ref{fig:Yamp_kappamod}, but including $\alpha$-heating. A value of $\kappa_{eff}=0.3$ was used, representing the most magnetized cases simulated (e.g. in figure \ref{fig:Ti_xMHD}). The yield was found to be amplified by a factor of 2 due to thermal conductivity suppression, which exceeds the expectations from the no-$\alpha$ theory displayed in figure \ref{fig:Yamp_theory} for hot-spots with high unmagnetized temperatures. These results broadly agree with more detailed 2-D simulations of high yield implosions \cite{zimmerman2021}. It is expected that once the perturbations are also accounted for, the benefit of the magnetic field will increase.
	
	In addition to the performance improvements, an applied magnetic field is expected to result in greater reproducibility and allow for higher convergence designs with larger 1-D yields.

	\section*{Acknowledgements}
	This work was performed under the auspices of the U.S. Department of Energy by Lawrence Livermore National Laboratory under Contract DE-AC52-07NA27344 and by the LLNL-LDRD program under Project Number 20-SI-002. 
	
	This document was prepared as an account of work sponsored by an agency of the United States government. Neither the United States government nor Lawrence Livermore National Security, LLC, nor any of their employees makes any warranty, expressed or implied, or assumes any legal liability or responsibility for the accuracy, completeness, or usefulness of any information, apparatus, product, or process disclosed, or represents that its use would not infringe privately owned rights. Reference herein to any specific commercial product, process, or service by trade name, trademark, manufacturer, or otherwise does not necessarily constitute or imply its endorsement, recommendation, or favoring by the United States government or Lawrence Livermore National Security, LLC. The views and opinions of authors expressed herein do not necessarily state or reflect those of the United States government or Lawrence Livermore National Security, LLC, and shall not be used for advertising or product endorsement purposes.
		
	\section*{References}
	
	\bibliographystyle{unsrt}

\begin{thebibliography}{10}
		
		\bibitem{gotchev2009}
		O.~V. Gotchev, P.~Y. Chang, J.~P. Knauer, D.~D. Meyerhofer, O.~Polomarov,
		J.~Frenje, C.~K. Li, M.~J.-E. Manuel, R.~D. Petrasso, J.~R. Rygg, F.~H.
		S{\'e}guin, and R.~Betti.
		\newblock Laser-{{Driven Magnetic-Flux Compression}} in {{High-Energy-Density
				Plasmas}}.
		\newblock {\em Physical Review Letters}, 103(21):215004, November 2009.
		
		\bibitem{knauer2010}
		J.~P. Knauer, O.~V. Gotchev, P.~Y. Chang, D.~D. Meyerhofer, O.~Polomarov,
		R.~Betti, J.~a.~Frenje, C.~K. Li, M.~J.-E. Manuel, R.~D. Petrasso, J.~R.
		Rygg, and F.~H. S{\'e}guin.
		\newblock Compressing magnetic fields with high-energy lasers.
		\newblock {\em Physics of Plasmas}, 17(056318), May 2010.
		
		\bibitem{walsh2019}
		C~A Walsh, K~Mcglinchey, J~K Tong, B~D Appelbe, A~Crilly, M~Zhang, and J~P
		Chittenden.
		\newblock Perturbation {{Modifications}} by {{Pre-magnetisation}} in {{Inertial
				Confinement Fusion Implosions}}.
		\newblock {\em Physics of Plasmas}, 096(December):1--12, 2019.
		
		\bibitem{perkins2017}
		L.~J. Perkins, D.~D.-M Ho, B.~G. Logan, G.~B. Zimmerman, M.~A. Rhodes, D.~J.
		Strozzi, D.~T. Blackfield, and S.~A. Hawkins.
		\newblock The potential of imposed magnetic fields for enhancing ignition
		probability and fusion energy yield in indirect-drive inertial confinement
		fusion.
		\newblock {\em Physics of Plasmas}, 24(6):062708--062708, June 2017.
		
		\bibitem{perkins2013}
		L.~J. Perkins, B.~G. Logan, G.~B. Zimmerman, and C.~J. Werner.
		\newblock Two-dimensional simulations of thermonuclear burn in ignition-scale
		inertial confinement fusion targets under compressed axial magnetic fields.
		\newblock In {\em Physics of {{Plasmas}}}, volume~20, pages 0--6, 2013.
		
		\bibitem{chandrasekhar1962}
		S.~Chandrasekhar and J.~Gillis.
		\newblock {\em Hydrodynamic and {{Hydromagnetic Stability}}}, volume~15.
		\newblock March 1962.
		
		\bibitem{srinivasan2013}
		Bhuvana Srinivasan and Xian-Zhu Tang.
		\newblock The mitigating effect of magnetic fields on {{Rayleigh-Taylor}}
		unstable inertial confinement fusion plasmas.
		\newblock {\em Physics of Plasmas}, 20(5):056307--056307, 2013.
		
		\bibitem{PhysRevE.104.L023201}
		J.~Narkis, F.~Conti, A.~L. Velikovich, and F.~N. Beg.
		\newblock Mitigation of magneto-{{Rayleigh-Taylor}} instability growth in a
		triple-nozzle, neutron-producing gas-puff {{Z}} pinch.
		\newblock {\em Physical Review E: Statistical Physics, Plasmas, Fluids, and
			Related Interdisciplinary Topics}, 104(2):L023201, August 2021.
		
		\bibitem{walsh2021magnetized}
		Christopher~A. Walsh.
		\newblock Magnetized ablative rayleigh-taylor instability in 3-{{D}}, 2021.
		
		\bibitem{walsh2020a}
		C.A. Walsh, A.J. Crilly, and J.P. Chittenden.
		\newblock Magnetized directly-driven {{ICF}} capsules: Increased instability
		growth from non-uniform laser drive.
		\newblock {\em Nuclear Fusion}, 60(10):106006, September 2020.
		
		\bibitem{matsuo2017}
		Kazuki Matsuo, Hideo Nagatomo, Zhe Zhang, Philippe Nicolai, Takayoshi Sano,
		Shohei Sakata, Sadaoki Kojima, Seung~Ho Lee, King Fai~Farley Law, Yasunobu
		Arikawa, Youichi Sakawa, Taichi Morita, Yasuhiro Kuramitsu, Shinsuke Fujioka,
		and Hiroshi Azechi.
		\newblock Magnetohydrodynamics of laser-produced high-energy-density plasma in
		a strong external magnetic field.
		\newblock {\em Physical Review E}, 95(5), 2017.
		
		\bibitem{epperlein1986}
		E.~M. Epperlein and M.~G. Haines.
		\newblock Plasma transport coefficients in a magnetic field by direct numerical
		solution of the {{Fokker}}\textendash{{Planck}} equation.
		\newblock {\em Physics of Fluids}, 29(4), 1986.
		
		\bibitem{davies2015}
		J.~R. Davies, R.~Betti, P.-Y. Chang, and G.~Fiksel.
		\newblock The importance of electrothermal terms in {{Ohm}}'s law for
		magnetized spherical implosions.
		\newblock {\em Physics of Plasmas}, 22(112703), November 2015.
		
		\bibitem{davies2017}
		J.~R. Davies, D.~H. Barnak, R.~Betti, E.~M. Campbell, P.-Y. Chang, A.~B.
		Sefkow, K.~J. Peterson, D.~B. Sinars, and M.~R. Weis.
		\newblock Laser-driven magnetized liner inertial fusion.
		\newblock {\em Physics of Plasmas}, 24(6):062701, June 2017.
		
		\bibitem{slutz2010}
		S.~A. Slutz, M.~C. Herrmann, R.~A. Vesey, A.~B. Sefkow, D.~B. Sinars, D.~C.
		Rovang, K.~J. Peterson, and M.~E. Cuneo.
		\newblock Pulsed-power-driven cylindrical liner implosions of laser preheated
		fuel magnetized with an axial field.
		\newblock {\em Physics of Plasmas}, 17(5):056303, March 2010.
		
		\bibitem{slutz2012}
		Stephen~A. Slutz and Roger~A. Vesey.
		\newblock High-{{Gain Magnetized Inertial Fusion}}.
		\newblock {\em Physical Review Letters}, 108(2):025003--025003, January 2012.
		
		\bibitem{slutz2016}
		S.~A. Slutz, W.~A. Stygar, M.~R. Gomez, K.~J. Peterson, A.~B. Sefkow, D.~B.
		Sinars, R.~A. Vesey, E.~M. Campbell, and R.~Betti.
		\newblock Scaling magnetized liner inertial fusion on {{Z}} and future
		pulsed-power accelerators.
		\newblock {\em Physics of Plasmas}, 23(2):022702, February 2016.
		
		\bibitem{slutz2018}
		S.~A. Slutz, M.~R. Gomez, S.~B. Hansen, E.~C. Harding, B.~T. Hutsel, P.~F.
		Knapp, D.~C. Lamppa, T.~J. Awe, D.~J. Ampleford, D.~E. Bliss, G.~A. Chandler,
		M.~E. Cuneo, M.~Geissel, M.~E. Glinsky, A.~J. {Harvey-Thompson}, M.~H. Hess,
		C.~A. Jennings, B.~Jones, G.~R. Laity, M.~R. Martin, K.~J. Peterson, J.~L.
		Porter, P.~K. Rambo, G.~A. Rochau, C.~L. Ruiz, M.~E. Savage, J.~Schwarz,
		P.~F. Schmit, G.~Shipley, D.~B. Sinars, I.~C. Smith, R.~A. Vesey, and M.~R.
		Weis.
		\newblock Enhancing performance of magnetized liner inertial fusion at the
		{{Z}} facility.
		\newblock {\em Physics of Plasmas}, 25(11):112706, November 2018.
		
		\bibitem{johzaki2016}
		T~Johzaki, H~Nagatomo, A~Sunahara, Y~Sentoku, H~Sakagami, M~Hata, T~Taguchi,
		K~Mima, Y~Kai, D~Ajimi, T~Isoda, T~Endo, A~Yogo, Y~Arikawa, S~Fujioka,
		H~Shiraga, and H~Azechi.
		\newblock Integrated simulation of magnetic-field-assist fast ignition laser
		fusion.
		\newblock {\em Plasma Physics and Controlled Fusion}, 59(1):014045, November
		2016.
		
		\bibitem{2021}
		C.A. Walsh, J.D. Sadler, and J.R. Davies.
		\newblock Updated magnetized transport coefficients: Impact on laser-plasmas
		with self-generated or applied magnetic fields.
		\newblock 61(11):116025, September 2021.
		
		\bibitem{barnak2017}
		D.~H. Barnak, J.~R. Davies, R.~Betti, M.~J. Bonino, E.~M. Campbell, V.~Yu.
		Glebov, D.~R. Harding, J.~P. Knauer, S.~P. Regan, A.~B. Sefkow, A.~J.
		{Harvey-Thompson}, K.~J. Peterson, D.~B. Sinars, S.~A. Slutz, M.~R. Weis, and
		P.-Y. Chang.
		\newblock Laser-driven magnetized liner inertial fusion on {{OMEGA}}.
		\newblock {\em Physics of Plasmas}, 24(5):056310, May 2017.
		
		\bibitem{hansen2020}
		E.~C. Hansen, J.~R. Davies, D.~H. Barnak, R.~Betti, E.~M. Campbell, V.~Yu.
		Glebov, J.~P. Knauer, L.~S. Leal, J.~L. Peebles, A.~B. Sefkow, and K.~M. Woo.
		\newblock Neutron yield enhancement and suppression by magnetization in
		laser-driven cylindrical implosions.
		\newblock {\em Physics of Plasmas}, 27(6):062703, June 2020.
		
		\bibitem{chang2011}
		P~Y Chang, G~Fiksel, M~Hohenberger, J~P Knauer, R~Betti, F~J Marshall, and D~D
		Meyerhofer.
		\newblock Fusion {{Yield Enhancement}} in {{Magnetized Laser-Driven
				Implosions}}.
		\newblock {\em Physical Review Letters}, 035006(July):2--5, 2011.
		
		\bibitem{ho2016}
		D.~D-M. Ho, L.~J. Perkins, G.~B. Zimmerman, B.~G. Logan, G.~Kagan, M.~A.
		Rhodes, and J.~D. Salmonson.
		\newblock In {\em {{EPS}}}, 2016.
		
		\bibitem{2018}
		O~A Hurricane, D~A Callahan, P~T Springer, M~J Edwards, P~Patel, K~Baker, D~T
		Casey, L~Divol, T~D{\"o}ppner, D~E Hinkel, L~F~Berzak Hopkins, A~Kritcher,
		S~Le Pape, S~Maclaren, L~Masse, A~Pak, L~Pickworth, J~Ralph, C~Thomas, A~Yi,
		and A~Zylstra.
		\newblock Beyond alpha-heating: Driving inertially confined fusion implosions
		toward a burning-plasma state on the {{National Ignition Facility}}.
		\newblock 61(1):014033, November 2018.
		
		\bibitem{spitzer1953}
		Lyman Spitzer and Richard H{\"a}rm.
		\newblock Transport {{Phenomena}} in a {{Completely Ionized Gas}}.
		\newblock {\em Physical Review}, 89(5):977--981, March 1953.
		
		\bibitem{ciardi2007}
		A.~Ciardi, S.~V. Lebedev, A.~Frank, E.~G. Blackman, J.~P. Chittenden, C.~J.
		Jennings, D.~J. Ampleford, S.~N. Bland, S.~C. Bott, J.~Rapley, G.~N. Hall,
		F.~A. {Suzuki-Vidal}, A.~Marocchino, T.~Lery, and C.~Stehle.
		\newblock The evolution of magnetic tower jets in the laboratory.
		\newblock In {\em Physics of {{Plasmas}}}, volume~14, November 2007.
		
		\bibitem{chittenden2009}
		J.P. Chittenden, N.P. Niasse, S.N. Bland, G.A. Hall, S.V. Lebedev, and C.~A.
		Jennings.
		\newblock Recent advances in magneto-hydrodynamic modeling of wire array
		{{Z-pinches}}.
		\newblock In {\em 2009 {{IEEE International Conference}} on {{Plasma Science}}
			- {{Abstracts}}}, pages 1--1. {IEEE}, June 2009.
		
		\bibitem{walsh2017}
		C.A. Walsh, J.P. Chittenden, K.~McGlinchey, N.P.L. Niasse, and B.D. Appelbe.
		\newblock Self-{{Generated Magnetic Fields}} in the {{Stagnation Phase}} of
		{{Indirect-Drive Implosions}} on the {{National Ignition Facility}}.
		\newblock {\em Physical Review Letters}, 118(15):155001--155001, April 2017.
		
		\bibitem{sharma2007}
		Prateek Sharma and Gregory~W. Hammett.
		\newblock Preserving monotonicity in anisotropic diffusion.
		\newblock {\em Journal of Computational Physics}, 227(1), November 2007.
		
		\bibitem{walsh2018a}
		C.~A. Walsh.
		\newblock {\em Extended {{Magneto-hydrodynamic Effects}} in {{Indirect-Drive
					Inertial Confinement Fusion Experiments}}}.
		\newblock PhD thesis, 2018.
		
		\bibitem{walsh2020}
		C.~A. Walsh, J.~P. Chittenden, D.~W. Hill, and C.~Ridgers.
		\newblock Extended-magnetohydrodynamics in under-dense plasmas.
		\newblock {\em Physics of Plasmas}, 27(2):022103, February 2020.
		
		\bibitem{walsh2021a}
		C.~A. Walsh and D.~S. Clark.
		\newblock Biermann battery magnetic fields in {{ICF}} capsules: {{Total}}
		magnetic flux generation.
		\newblock {\em Physics of Plasmas}, 28(9):092705, September 2021.
		
		\bibitem{PhysRevLett.125.145001}
		P.~T. Campbell, C.~A. Walsh, B.~K. Russell, J.~P. Chittenden, A.~Crilly,
		G.~Fiksel, P.~M. Nilson, A.~G.~R. Thomas, K.~Krushelnick, and L.~Willingale.
		\newblock Magnetic signatures of radiation-driven double ablation fronts.
		\newblock {\em Physical Review Letters}, 125(14):145001, September 2020.
		
		\bibitem{campbell2021measuring}
		P.~T. Campbell, C.~A. Walsh, B.~K. Russell, J.~P. Chittenden, A.~Crilly,
		G.~Fiksel, L.~Gao, I.~V. Igumenshchev, P.~M. Nilson, A.~G.~R. Thomas,
		K.~Krushelnick, and L.~Willingale.
		\newblock Measuring magnetic flux suppression in high-power laser-plasma
		interactions, 2021.
		
		\bibitem{sadler2021}
		James~D. Sadler, Christopher~A. Walsh, and Hui Li.
		\newblock Symmetric {{Set}} of {{Transport Coefficients}} for {{Collisional
				Magnetized Plasma}}.
		\newblock {\em Physical Review Letters}, 126(7):075001, February 2021.
		
		\bibitem{davies2021}
		J.~R. Davies, H.~Wen, Jeong-Young Ji, and Eric~D. Held.
		\newblock Transport coefficients for magnetic-field evolution in inviscid
		magnetohydrodynamics.
		\newblock {\em Physics of Plasmas}, 28(1):012305, January 2021.
		
		\bibitem{10.1088/1361-6587/ac3f25}
		Chris Walsh, Ricardo Florido, Mathieu {Bailly-Grandvaux}, Francisco
		{Suzuki-Vidal}, Jeremy~P Chittenden, Aidan Crilly, Marco~A Gigosos, Roberto
		Mancini, Gabriel {Perez-Callejo}, Christos Vlachos, Christopher McGuffey,
		Farhat~N Beg, and Joao~Jorge Santos.
		\newblock Exploring extreme magnetization phenomena in directly-driven
		imploding cylindrical targets.
		\newblock {\em Plasma Physics and Controlled Fusion}, 2021.
		
		\bibitem{clark2019}
		D.~S. Clark, C.~R. Weber, J.~L. Milovich, A.~E. Pak, D.~T. Casey, B.~A. Hammel,
		D.~D. Ho, O.~S. Jones, J.~M. Koning, A.~L. Kritcher, M.~M. Marinak, L.~P.
		Masse, D.~H. Munro, M.~V. Patel, P.~K. Patel, H.~F. Robey, C.~R. Schroeder,
		S.~M. Sepke, and M.~J. Edwards.
		\newblock Three-dimensional modeling and hydrodynamic scaling of {{National
				Ignition Facility}} implosions.
		\newblock {\em Physics of Plasmas}, 26(5):050601, 2019.
		
		\bibitem{Tong_2019}
		J.K. Tong, K.~McGlinchey, B.D. Appelbe, C.A. Walsh, A.J. Crilly, and J.P.
		Chittenden.
		\newblock Burn regimes in the hydrodynamic scaling of perturbed inertial
		confinement fusion hotspots.
		\newblock {\em Nuclear Fusion}, 59(8):086015, June 2019.
		
		\bibitem{appelbe2021}
		B.~Appelbe, A.~L. Velikovich, M.~Sherlock, C.~Walsh, A.~Crilly, S.~O'~Neill,
		and J.~Chittenden.
		\newblock Magnetic field transport in propagating thermonuclear burn.
		\newblock {\em Physics of Plasmas}, 28(3):032705, March 2021.
		
		\bibitem{casey2018}
		D.~T. Casey, C.~A. Thomas, K.~L. Baker, B.~K. Spears, M.~Hohenberger, S.~F.
		Khan, R.~C. Nora, C.~R. Weber, D.~T. Woods, O.~A. Hurricane, D.~A. Callahan,
		R.~L. Berger, J.~L. Milovich, P.~K. Patel, T.~Ma, A.~Pak, L.~R. Benedetti,
		M.~Millot, C.~Jarrott, O.~L. Landen, R.~M. Bionta, B.~J. MacGowan, D.~J.
		Strozzi, M.~Stadermann, J.~Biener, A.~Nikroo, C.~S. Goyon, N.~Izumi, S.~R.
		Nagel, B.~Bachmann, P.~L. Volegov, D.~N. Fittinghoff, G.~P. Grim, C.~B.
		Yeamans, M.~Gatu~Johnson, J.~A. Frenje, N.~Rice, C.~Kong, J.~Crippen,
		J.~Jaquez, K.~Kangas, and C.~Wild.
		\newblock The high velocity, high adiabat, ``{{Bigfoot}}'' campaign and tests
		of indirect-drive implosion scaling.
		\newblock {\em Physics of Plasmas}, 25(5):056308, May 2018.
		
		\bibitem{thomas2020}
		C.~A. Thomas, E.~M. Campbell, K.~L. Baker, D.~T. Casey, M.~Hohenberger, A.~L.
		Kritcher, B.~K. Spears, S.~F. Khan, R.~Nora, D.~T. Woods, J.~L. Milovich,
		R.~L. Berger, D.~Strozzi, D.~D. Ho, D.~Clark, B.~Bachmann, L.~R. Benedetti,
		R.~Bionta, P.~M. Celliers, D.~N. Fittinghoff, G.~Grim, R.~Hatarik, N.~Izumi,
		G.~Kyrala, T.~Ma, M.~Millot, S.~R. Nagel, P.~K. Patel, C.~Yeamans, A.~Nikroo,
		M.~Tabak, M.~Gatu~Johnson, P.~L. Volegov, and S.~M. Finnegan.
		\newblock Deficiencies in compression and yield in x-ray-driven implosions.
		\newblock {\em Physics of Plasmas}, 27(11):112705, November 2020.
		
		\bibitem{moody2021}
		J.~D. Moody.
		\newblock Boosting {{Inertial-Confinement-Fusion Yield}} with {{Magnetized
				Fuel}}.
		\newblock {\em Physics}, May 2021.
		
		\bibitem{moody2021a}
		J.~D. Moody.
		\newblock The magnetized indirect drive project on the {{National Ignition
				Facility}}.
		\newblock {\em In Submission}, 2021.
		
		\bibitem{strozzi2015}
		David~J. Strozzi, L.~J. Perkins, M.~M. Marinak, D.~J. Larson, J.~M. Koning, and
		B.~G. Logan.
		\newblock Imposed magnetic field and hot electron propagation in inertial
		fusion hohlraums.
		\newblock {\em Journal of Plasma Physics}, 81(06):475810603--475810603,
		December 2015.
		
		\bibitem{montgomery2015}
		D.~S. Montgomery, B.~J. Albright, D.~H. Barnak, P.~Y. Chang, J.~R. Davies,
		G.~Fiksel, D.~H. Froula, J.~L. Kline, M.~J. MacDonald, A.~B. Sefkow, L.~Yin,
		and R.~Betti.
		\newblock Use of external magnetic fields in hohlraum plasmas to improve
		laser-coupling.
		\newblock {\em Physics of Plasmas}, 22(1):010703--010703, January 2015.
		
		\bibitem{strozzi}
		D.~J. Strozzi, J.~D. Moody, B.~B. Pollock, H.~Sio, G.~Zimmerman, D.~Ho, S.~O.
		Kucheyev, C.~A. Walsh, and B.~G. Logan.
		\newblock First {{Magnetized Hohlraum-Driven Implosions}} on the {{NIF}}.
		\newblock In {\em {{APS DPP}} 2021}.
		
		\bibitem{PhysRevLett.100.185006}
		T.~C. Sangster, V.~N. Goncharov, P.~B. Radha, V.~A. Smalyuk, R.~Betti, R.~S.
		Craxton, J.~A. Delettrez, D.~H. Edgell, V.~Yu. Glebov, D.~R. Harding,
		D.~{Jacobs-Perkins}, J.~P. Knauer, F.~J. Marshall, R.~L. McCrory, P.~W.
		McKenty, D.~D. Meyerhofer, S.~P. Regan, W.~Seka, R.~W. Short, S.~Skupsky,
		J.~M. Soures, C.~Stoeckl, B.~Yaakobi, D.~Shvarts, J.~A. Frenje, C.~K. Li,
		R.~D. Petrasso, and F.~H. S{\'e}guin.
		\newblock High-areal-density fuel assembly in direct-drive cryogenic
		implosions.
		\newblock {\em Physical Review Letters}, 100(18):185006, May 2008.
		
		\bibitem{crilly2020}
		A.~J. Crilly, B.~D. Appelbe, O.~M. Mannion, C.~J. Forrest, V.~Gopalaswamy,
		C.~A. Walsh, and J.~P. Chittenden.
		\newblock Neutron backscatter edge: {{A}} measure of the hydrodynamic
		properties of the dense {{DT}} fuel at stagnation in {{ICF}} experiments.
		\newblock {\em Physics of Plasmas}, 27(1):012701, January 2020.
		
		\bibitem{PhysRevLett.127.165001}
		Kazuki Matsuo, Takayoshi Sano, Hideo Nagatomo, Toshihiro Somekawa, King
		Fai~Farley Law, Hiroki Morita, Yasunobu Arikawa, and Shinsuke Fujioka.
		\newblock Enhancement of ablative rayleigh-taylor instability growth by thermal
		conduction suppression in a magnetic field.
		\newblock {\em Physical Review Letters}, 127(16):165001, October 2021.
		
		\bibitem{bose2021}
		A~Bose.
		\newblock Effect of strongly magnetized electrons and ions on heat flow and
		symmetry of intertial fusion implosions.
		\newblock {\em In Submission}, 2021.
		
		\bibitem{betti2001}
		R.~Betti, M.~Umansky, V.~Lobatchev, V.~N. Goncharov, and R.~L. McCrory.
		\newblock Hot-spot dynamics and deceleration-phase
		{{Rayleigh}}\textendash{{Taylor}} instability of imploding inertial
		confinement fusion capsules.
		\newblock {\em Physics of Plasmas}, 8(12):5257--5267, December 2001.
		
		\bibitem{lyon1995}
		S~P Lyon and J~D Johnson.
		\newblock Sesame: {{The Los Alamos National Laboratory}} equation of {{State
				Database}}.
		\newblock Technical {{Report}}, {Los Alamos National Laboratory}, 1995.
		
		\bibitem{merrill2012}
		F.~E. Merrill, D.~Bower, R.~Buckles, D.~D. Clark, C.~R. Danly, O.~B. Drury,
		J.~M. Dzenitis, V.~E. Fatherley, D.~N. Fittinghoff, R.~Gallegos, G.~P. Grim,
		N.~Guler, E.~N. Loomis, S.~Lutz, R.~M. Malone, D.~D. Martinson, D.~Mares,
		D.~J. Morley, G.~L. Morgan, J.~A. Oertel, I.~L. Tregillis, P.~L. Volegov,
		P.~B. Weiss, C.~H. Wilde, and D.~C. Wilson.
		\newblock The neutron imaging diagnostic at {{NIF}} (invited).
		\newblock {\em Review of Scientific Instruments}, 83(10):10D317, October 2012.
		
		\bibitem{mcglinchey2018}
		Kristopher McGlinchey, Brian Appelbe, Aidan Crilly, Jon Tong, Christopher
		Walsh, and Jeremy Chittenden.
		\newblock Diagnostic {{Signatures}} of {{Performance Degrading Perturbations}}
		in {{Inertial Confinement Fusion Implosions}}.
		\newblock {\em Physics of Plasmas, Accepted}, 2018.
		
		\bibitem{1993}
		H.-S Bosch and G.M Hale.
		\newblock Improved formulas for fusion cross-sections and thermal reactivities.
		\newblock 33(12):1919--1919, December 1993.
		
		\bibitem{huba2013}
		J.~B. Huba.
		\newblock {{NRL FORMULARY}}, 2013.
		
		\bibitem{PhysRevLett.102.025004}
		P.~Michel, L.~Divol, E.~A. Williams, S.~Weber, C.~A. Thomas, D.~A. Callahan,
		S.~W. Haan, J.~D. Salmonson, S.~Dixit, D.~E. Hinkel, M.~J. Edwards, B.~J.
		MacGowan, J.~D. Lindl, S.~H. Glenzer, and L.~J. Suter.
		\newblock Tuning the implosion symmetry of {{ICF}} targets via controlled
		crossed-beam energy transfer.
		\newblock {\em Physical Review Letters}, 102(2):025004, January 2009.
		
		\bibitem{PhysRevLett.111.235001}
		E.~L. Dewald, J.~L. Milovich, P.~Michel, O.~L. Landen, J.~L. Kline, S.~Glenn,
		O.~Jones, D.~H. Kalantar, A.~Pak, H.~F. Robey, G.~A. Kyrala, L.~Divol, L.~R.
		Benedetti, J.~Holder, K.~Widmann, A.~Moore, M.~B. Schneider, T.~D{\"o}ppner,
		R.~Tommasini, D.~K. Bradley, P.~Bell, B.~Ehrlich, C.~A. Thomas, M.~Shaw,
		C.~Widmayer, D.~A. Callahan, N.~B. Meezan, R.~P.~J. Town, A.~Hamza,
		B.~Dzenitis, A.~Nikroo, K.~Moreno, B.~Van~Wonterghem, A.~J. Mackinnon, S.~H.
		Glenzer, B.~J. MacGowan, J.~D. Kilkenny, M.~J. Edwards, L.~J. Atherton, and
		E.~I. Moses.
		\newblock Early-time symmetry tuning in the presence of cross-beam energy
		transfer in {{ICF}} experiments on the national ignition facility.
		\newblock {\em Physical Review Letters}, 111(23):235001, December 2013.
		
		\bibitem{PhysRevE.98.051201}
		M.~Gatu~Johnson, B.~D. Appelbe, J.~P. Chittenden, J.~Delettrez, C.~Forrest,
		J.~A. Frenje, V.~Yu. Glebov, W.~Grimble, B.~M. Haines, I.~Igumenshchev,
		R.~Janezic, J.~P. Knauer, B.~Lahmann, F.~J. Marshall, T.~Michel, F.~H.
		S{\'e}guin, C.~Stoeckl, C.~Walsh, A.~B. Zylstra, and R.~D. Petrasso.
		\newblock Impact of asymmetries on fuel performance in inertial confinement
		fusion.
		\newblock {\em Physical Review E: Statistical Physics, Plasmas, Fluids, and
			Related Interdisciplinary Topics}, 98(5):051201, November 2018.
		
		\bibitem{gatujohnson2019}
		M.~Gatu~Johnson, B.~D. Appelbe, J.~P. Chittenden, A.~Crilly, J.~Delettrez,
		C.~Forrest, J.~A. Frenje, V.~Yu. Glebov, W.~Grimble, B.~M. Haines, I.~V.
		Igumenshchev, R.~Janezic, J.~P. Knauer, B.~Lahmann, F.~J. Marshall,
		T.~Michel, F.~H. S{\'e}guin, C.~Stoeckl, C.~Walsh, A.~B. Zylstra, and R.~D.
		Petrasso.
		\newblock Impact of imposed mode 2 laser drive asymmetry on inertial
		confinement fusion implosions.
		\newblock {\em Physics of Plasmas}, 26(1):012706, January 2019.
		
		\bibitem{clark2016}
		D.~S. Clark, C.~R. Weber, J.~L. Milovich, J.~D. Salmonson, A.~L. Kritcher,
		S.~W. Haan, B.~A. Hammel, D.~E. Hinkel, O.~A. Hurricane, O.~S. Jones, M.~M.
		Marinak, P.~K. Patel, H.~F. Robey, S.~M. Sepke, and M.~J. Edwards.
		\newblock Three-dimensional simulations of low foot and high foot implosion
		experiments on the {{National Ignition Facility}}.
		\newblock {\em Physics of Plasmas}, 23(5):056302--056302, May 2016.
		
		\bibitem{casey2021}
		D.~T. Casey, B.~J. MacGowan, J.~D. Sater, A.~B. Zylstra, O.~L. Landen,
		J.~Milovich, O.~A. Hurricane, A.~L. Kritcher, M.~Hohenberger, K.~Baker,
		S.~Le~Pape, T.~D{\"o}ppner, C.~Weber, H.~Huang, C.~Kong, J.~Biener, C.~V.
		Young, S.~Haan, R.~C. Nora, S.~Ross, H.~Robey, M.~Stadermann, A.~Nikroo,
		D.~A. Callahan, R.~M. Bionta, K.~D. Hahn, A.~S. Moore, D.~Schlossberg,
		M.~Bruhn, K.~Sequoia, N.~Rice, M.~Farrell, and C.~Wild.
		\newblock Evidence of three-dimensional asymmetries seeded by high-density
		carbon-ablator nonuniformity in experiments at the national ignition
		facility.
		\newblock {\em Physical Review Letters}, 126(2):025002, January 2021.
		
		\bibitem{zimmerman2021}
		G.~B. Zimmerman, D.~Ho, A.~L. Velikovich, Russell~M Kulsrud, J.~D. Moody,
		J.~Harte, and A~L Kritcher.
		\newblock Magnetized {{ICF}}: Role of e-thermal conductivity on imploding shock
		and high-yield capsule designs.
		\newblock In {\em {{APS DPP}}}, 2021.
		
	\end{thebibliography}

\end{document}